\documentclass[3p,times]{elsarticle}

\usepackage{cuted}
\usepackage{amsmath}
\usepackage{amssymb}
\usepackage{fixltx2e}
\usepackage{stfloats}
\usepackage[figuresright]{rotating}

\usepackage{booktabs}
\usepackage{lscape}
\usepackage{longtable}
\usepackage{dcolumn}
\usepackage{graphicx}
\usepackage{subfig}

\begin{document}

\begin{frontmatter}

\title{Dynamical analysis of biological signals with the 0--1 test}

\author[UAH]{J. de Pedro-Carracedo\corref{cor1}}
\ead{javier.depedro@uah.es}
\author[UPM1]{A.M. Ugena}
\ead{anamaria.ugena@upm.es}
\author[UPM2]{A.P. Gonzalez-Marcos\corref{cor2}}
\ead{anapilar.gonzalez@upm.es}
\cortext[cor1]{Corresponding author}
\cortext[cor2]{Principal corresponding author}
\address[UAH]{University of Alcal\'a (UAH), Computer Engineering Department, Alcal\'a de Henares (Madrid), 28871 Spain}
\address[UPM1]{Technical University of Madrid (UPM), Departamento de Matem\'atica aplicada a las Tecnolog\'ias de la Informaci\'on, Madrid, 28040 Spain}
\address[UPM2]{Technical University of Madrid (UPM), Photonic Technology and Bioengineering Department, Madrid, 28040 Spain}

\begin{abstract}
The 0--1 test distinguishes between regular and chaotic dynamics for a deterministic system using a time series. Here is presented how to use the test with biological signals. Between all the biological signals obtained as a times series, we present the results of applying the test and study with more details the photoplethysmogram (PPG) from different subjects; different biological signals of one subject are analyzed also with the test. PPG signal contains extensive physiological dynamics information; mainly used to heart rate monitoring and blood oxygen saturation. Based on its understandable statistics, several indices have been proposed and may apply to some clinical needs. However, continuous real-time monitoring, of the non-stationary biological signals, is frequently used on the modern intensive care unit. A deep understanding of the PPG signal will allow finding algorithms to evaluate other possible indices that give information about the dynamical system that constitutes the human body. We demonstrate that the PPG signal, as a time series, on a healthy individual, is a quasi-periodic signal. The parameters calculated for the analysis give some ideas of what the quasi-periodic properties mean. Results are on PPG signals from different healthy young people. The dynamical analysis of the PPG signal applying the 0--1 test allows distinguishing a regular dynamic, periodic o quasi-periodic, from a chaotic dynamics operating directly over the time series. The result obtained is compared with the results from well known time series that are random, chaotic, aperiodic, periodic, and quasi-periodic. 
\end{abstract}


\end{frontmatter}

%\linenumbers

\section{Introduction}
The physiological system is limited to the field of dissipative dynamic systems since energy dissipation results extremely useful to stabilize biological dynamics: preserve proper functioning under optimal conditions of a living organism \cite{Seely2012}.

In the context of dissipative dynamic systems, recurrence properties and ergodicity facilitate the analysis of physical processes from an only scalar time series; which is nothing other than a sequence of measurements of an observable, regular procedure, or of several observables of a physical system, performed at regular intervals of time. In the case that concerns us, we study a sequence of measurements of an observable, the PPG signal. The PPG signal takes its name from the optical technique apply for volumetric analysis of an organ (plethysmogram), the photoplethysmogram; or from the device used to measure it, the photoplethysmograph, introduced by A. Hertzman in 1937 \cite{Hertzman1937}. This signal is also known as a peripheral pulse wave, mainly employed in clinical with a peripheral pulse monitor or pulse oximeter \cite{Murray1996}.   

Between the different dynamic behavior that can be present in the dynamical systems theory, the chaos is the behavior most complex and representative when the evolution is nonlinear. Chaos is a deterministic behavior. One way to extract information from a system is to analyze the deterministic degree of its behavior, being able to identify if the dynamic is regular, periodic, quasi-periodic, chaotic, or random. As far as we know, there are only a few studies that analyze a PPG signal from a nonlinear point of view, being representative \cite{Bhattacharya2001}, the stochastic model by \cite{Martin-Martinez2013}, and the work of Nina Sviridova \cite{Sviridova2018}.

In 2004, Gottwald and Melbourne proposed a new method to test chaos in a deterministic system, the 0--1 test \cite{Gottwald2004}. This test allows---calculating the grown rate $K_{c}$, with a value of 0 for periodic behavior, and a value of 1 for a chaotic one---obtain information of the type of evolution of a dynamical system without the requirement of phase space reconstruction from a real-world time series. We present that the 0--1 test allows to also evaluated the main characteristic of an underlying system from a time series. We have applied it to the PPG signals from several subjects. Results of applying 0--1 test to other biological signals, as electromyography (EMG) and, finger and cheek, temperature from just one subject is also presented to compare the PPG signal. Future publications will present a more detailed study of other biological signals. Data are from project \cite{Aguilo2015}.

The paper has four more sections. The theoretical underpinning of the 0--1 test is described briefly in Section \ref{sec:01TEST}. Section \ref{sec:ALGORITHM} presents the algorithm applied to obtain the results. In Section \ref{sec:SIGNALS}, we show the references signals used to compare the behavior of the PPG, followed by graphical results and discussion in Section \ref{sec:RESULTS}. Conclusions are in Section \ref{sec:CONCLUSION}.

\section{0--1 test description}\label{sec:01TEST}

The test, development in \cite{Gottwald2004, Gottwald2005, Gottwald2008}, uses as input a time series of discrete scalar data $s(n)$, for $n= 1,2,\dotsc,N$, where $s(n)$ is a $1$-dimension observable of the underlying dynamical system. An observable is everything that can be measured (the value of all the observables defines a state). The data used is directly from measurement including noise.  The test output is a bi-dimensional system: 
\begin{eqnarray}\label{eqn:SD01}
p_{c}(n) &=& \sum_{j=1}^{n}s(j)\cos jc,\nonumber \\
q_{c}(n) &=& \sum_{j=1}^{n}s(j)\sin jc,
\end{eqnarray}
where $c\in(0,2\pi)$ is a constant value, and $n=1,2,\dotsc,N$. The mean squared displacement (MSD) or mean square fluctuation, measures of the deviation of the position of a particle for a reference position over time, $M_{c}(n)$, is defined as:

\begin{equation}\label{eqn:SD02}
M_{c}(n)=\frac{1}{N}\sum_{j=1}^{N-n}\left([p_{c}(j+n)-p_{c}(j)]^{2}+[q_{c}(j+n)-q_{c}(j)]^{2}\right),\quad n=1,2,3,\dotsc
\end{equation}

%\begin{multline}\label{eqn:SD02}
%M_{c}(n)=\frac{1}{N}\sum_{j=1}^{N-n}\left([p_{c}(j+n)-
%p_{c}(j)]^{2}+\right.\\
%\left.[q_{c}(j+n)-q_{c}(j)]^{2}\right),\quad n=1,2,3,\dotsc
%\end{multline}

The MSD changes with time, as $n$ grows. The way to know how changes are giving by the growth rate, $K_{c}$:
\begin{equation}\label{eqn:SD03}
K_{c}=\underset{n\rightarrow \infty}{\lim}\frac{\log M_{c}(n)}{\log n}.
\end{equation}

The values of $K_{c}$ determine the type of behavior of a dynamic system. If $K_{c}\approx 0$, the dynamic system behaves on a regularly; if $K_{c}\approx 1$ evolves chaotically. In the case of regular dynamic systems evolutions, defined by the equations of the system \eqref{eqn:SD01}, describe bounded paths; in chaotic dynamical systems the evolution experienced by the paths proposed at \eqref{eqn:SD01} resembles a $2$-dimensional Brownian motion, that is, its evolution is diffusive.

  The average quadratic displacement $M_{c}(n)$ enables the identification of the growth rate of the paths, whether it be a bounded growth rate (regular case) or a rate of linear growth (if chaotic). In any case, the parameter $K_{c}$ captures this growth rate.

\section{0--1 test algorithm}\label{sec:ALGORITHM}
To synthesize the 0--1 test process \cite{Skokos2016}, Table \ref{tab:T01} shows the four steps to follow. The actions of the algorithm would be:
\begin{enumerate}
\item Equation \eqref{eqn:SD01} is solved for $c\in (0,2\pi)$. 
\item To analyze the diffusive or non-diffusive behavior, equation \eqref{eqn:SD02} is computed with $p_{c}(n)$ and $q_{c}(n)$.

Gottwald and Melbourne \cite{Gottwald2009} proposes an expression similar to \eqref{eqn:SD02}, regarding asymptotic behavior, with best properties of convergence. 

\item The rate of asymptotic growth $K_{c}$ of $M_{c}(n)$, allows differentiating between regular behavior and chaotic behavior. Two methods used to calculate $K_{c}$: regression method and correlation method.
The correlation method provides better results than the regression method, according to the analyses carried out, for different dynamic systems, in \cite{Gottwald2009}. In our study, we have used both types of methods, and we have verified that indeed, the correlation method yields better results. 
\item Run steps 1 to 3 for several values of $c$; with 100 $c$ values, selected randomly, it is enough. The final result, $K_{m}$, attends to the mean of the computed $K_ {c}$ values, that is, $K_{m}=\text{mean}(K_{c})$.\footnote{In this study we introduce a slight modification to the original method, taking the absolute value of the calculated $K_{c}$ values, since the interest is focused not so much on the numerical values as on the strength of the correlation.}
\end{enumerate}

\begin{table*}[ht]
\renewcommand{\arraystretch}{2.3}
\caption{Steps of 0--1 test algorithm.}
\label{tab:T01}
\centering
\begin{tabular}{|l||l||c||l|}
\hline
\begin{minipage}[c][1cm][c]{2.75cm}
Step 1: From equation \eqref{eqn:SD01}
\end{minipage} &
\begin{minipage} [c][1cm][c]{6cm}
Resolve: $p_{c}(n)$ and $q_{c}(n)$, $n= 1,2,\dotsc,N$;\\ $c\in(0,2\pi)$
\end{minipage}
& \begin{minipage}[c][1cm][c]{2.1cm}
Number of observations: $N$
\end{minipage} &  \begin{minipage}[c][1cm][c]{2.9cm}
 Plots: Fig. \ref{fig:F01}(f)-(j),\\Fig. \ref{fig:F02}(f)-(j)
\end{minipage}\\
%From equation (1) &pc(n) y qc(n) n= 1,2‰??,N;
%c (0,ì?)& N  & \\
\hline
\hline
\begin{minipage}[c][2.45cm][c]{2.75cm}
Step 2: Analyses the diffusive, or non-diffusive, behavior of 
$p_{c}(n)$ and $q_{c}(n)$
\end{minipage} &
\begin{minipage}[c][2cm][c]{5cm}
%\centering
\begin{flushleft}
\begin{multline*}
M_{c}(n)=\frac{1}{N}\sum_{j=1}^{N-n}\left([p_{c}(j+n)-p_{c}(j)]^{2}+\right.\\
\left.[q_{c}(j+n)-q_{c}(j)]^{2}\right),\quad n=1,2,3,\dotsc
\end{multline*}
\end{flushleft}
\end{minipage}
& \begin{minipage}[c][1cm][c]{2.1cm}
\centering
$n\leqslant N_{0} \ll N$
\end{minipage} &  \begin{minipage}[c][1cm][c]{2.9cm}
 The plot of this step is not relevant in this work
\end{minipage}\\
\hline
\hline
\begin{minipage}[c][1cm][c]{2.75cm}
Step 3: Grown rate $K_{c}$
\end{minipage} &
\multicolumn{3}{>{}c|}{Regression or correlation method}\\

\hline
\hline
\begin{minipage}[c][2.25cm][c]{2.75cm}
Step 4: 
Steps 1--3 must be executed for various values of $c$ (randomly selected)
\end{minipage} &
\multicolumn{2}{>{}c||}{\begin{minipage} [c][1cm][c]{9.2cm}In practice 100 choices of $c\in (0,2\pi)$ are sufficient; moreover, in our case, $K_{m}=\text{mean}(K_{c})$ about the central region of the calculated $K_{c}$ values\end{minipage}}
&  \begin{minipage}[c][1cm][c]{2.9cm}
 Plots: Fig. \ref{fig:F01}(k)-(o),\\Fig. \ref{fig:F02}(k)-(o)
\end{minipage}\\
\hline
\end{tabular}
\end{table*}

In any of the methods used a $K_{m}\approx 0$ value indicates a regular dynamic, and a value of $K_{m}\approx 1$ suggests a chaotic behavior.

 Commonly, a length of long enough time series is claimed in any chaos detection method to make the attractor revealed as a whole, as the accuracy of the tests depends on the asymptotic behavior of the underlying system, whether diffusive or not.

If the length of the time series under study is short, it is within the possibility that the asymptotic behavior, diffusive or not, lacks the dominant effect expected. In this case, more than the convergence of $K_{c}$ towards 0 or 1, the tendency of $K_{c}$ towards 0 or 1, concerning the length of the time series, is examined to dictate the presence or absence of chaos.

\section{Reference signals and PPG}\label{sec:SIGNALS}
Although the complexity of signals, including the acquisition system, admits different definitions, a consistent alternative evaluates the degree of complexity regarding regularity in the patterns of repetition of the data. These lead to an ordering of the signals between two opposite ends, the most regular or periodic and the most irregular or random, with a whole range of intermediate options; as they approach a random regime, as in the case of chaotic signals, acquire greater degrees of freedom or versatility, without losing determinism in its dynamic behavior.  

We apply the 0--1 test to different signals that represent the typical behavior of time series concerning regularity in the patterns of repetition of the data.  The classification of the signals is according to the order of less to greater complexity, and Fig. \ref{fig:F01}(a)-(e) represent the Power Spectrum Density (PSD) of all of them normalized in amplitude to the interval $[0, 1]$.

\subsection*{Periodic}
A repetitive pattern occurs and a Fourier series can describe the signal. If a sine wave is analyzed the value of $K_{m}\approx 0.18$; to study a more complex signal, with two frequencies with a rational relation, we use saw's wave at $f=100$ Hz:

\begin{equation}\label{eqn:SP}
y_{\text{periodic}}(t)=2\left(\frac{t}{1/f}-\left\lfloor \frac{1}{2}+\frac{t}{1/f}\right\rfloor\right).
\end{equation}

\subsection*{Quasi-periodic}
A Fourier series can also decompose it. The relation between frequencies is irrational \cite{Landau1987}, forming a torus or tori depending on the grades of freedom. In this case, we apply the most simple one, two cosine waves with an irrational relation $1/\sqrt{2}$:

\begin{equation}\label{eqn:SQ}
y_{\text{quasi-periodic}}(t)=\cos\left(2\pi 100\cdot t\right)+\cos\left(2\pi100\sqrt{2}\cdot t \right).
\end{equation}

\subsection*{Aperiodic} 
Aperiodic has non self-similar repetition. Similar to a periodic function with an infinite period. The function choose is the one that generates samples, by a linear frequency sweeping the \textit{chirp} function: $f(t)=f_{0}+k\cdot t$, between $f_{0}=0$ Hz and $f_{1}=100$ Hz, with $k=(f_{1}-f_{0})/T$, being $T$ the sweep time; and the sampling frequency $f_{s}=5$ kHz.

\begin{equation}\label{eqn:SA}
y_{\text{aperiodic}}(t)=\sin\left[2\pi\left(f_{0}t+\frac{k}{2}t^{2}\right)\right].
\end{equation}

\subsection*{Chaotic}
The sensitivity of the deterministic function to small changes in the initial state is the characteristic footprint of a chaotic behaviour. The initial uncertainty increases with time, and it is not possible to predict the final state of the system (N. S. Krylov 1944; M. Born 1952) \cite{Landau1987}. We use the H\'enon map dynamic variable $x_{n}$, with $a=1.4$ y $b=0.3$. The initial conditions were $x_ {0}= y_{0}= 0.03$. 100,000 samples were generated, using the last 5,000 points. 

\begin{eqnarray}\label{eqn:SC}
x_{n+1} &=& 1-ax_{n}^{2}+y_{n},\nonumber \\
y_{n+1} &=& bx_{n}.
\end{eqnarray}

\subsection*{Random} 
It is not deterministic and requires a probabilistic characterization. 5,000 points were generated by a uniform distribution in the interval $[0,1]$.
\subsection*{Biological signals}
\paragraph{PPG signal}
A PPG signal has several components that include information about: volumetric changes in the arterial blood associated with the cardiac activity; variations in venous blood volume which modulate the PPG signal; the tissues• optical property and subtle energy changes in the body which determine the DC component. A good review is in \cite{Allen2007}.

These means that PPG can be considered a signal print subject for a short period, and allows the evaluation of the evolution of a body. 

To be able to evaluate the information provided by this signal, we analyzed the dynamical system under it with nonlinear techniques. In this work, we used the finger PPG signals obtained from \cite{Aguilo2015}. This publication shows the data of six subjects and only two seconds of the total data recorded in the indicated project.  

\paragraph{Other signals}
Electromyography (EMG), measured in two different points, in trapezius muscle and face (under the eye), and, finger and cheek, temperature from just one subject are also presented to compare the PPG signal. All of them are study with the 0--1 test, results in the next section. Again, this publication shows only two seconds of the available data.

\begin{figure*}[!t]
\centering
\subfloat[]{\includegraphics[width=1.25in]{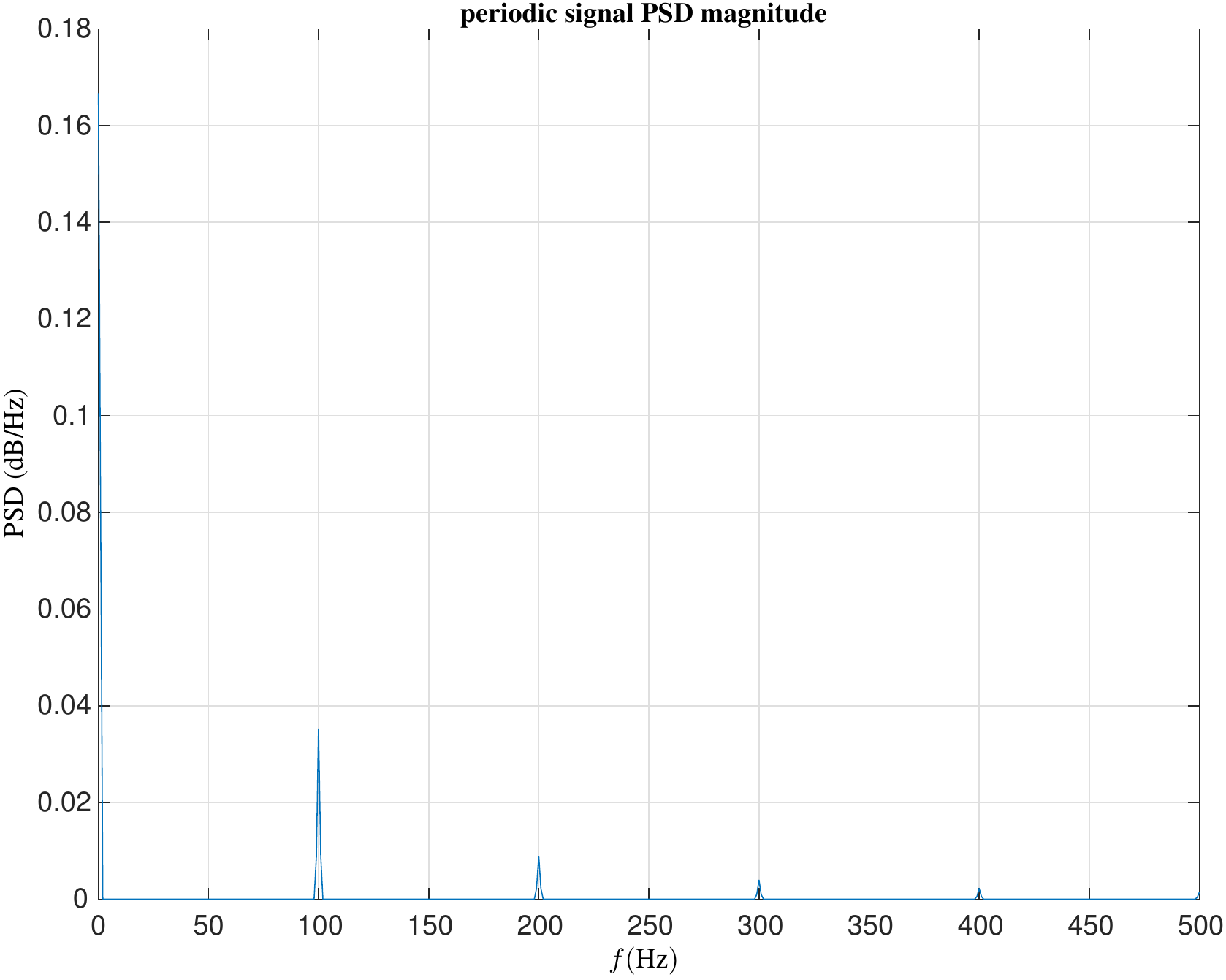}
\label{fig_first_case}}
\subfloat[]{\includegraphics[width=1.25in]{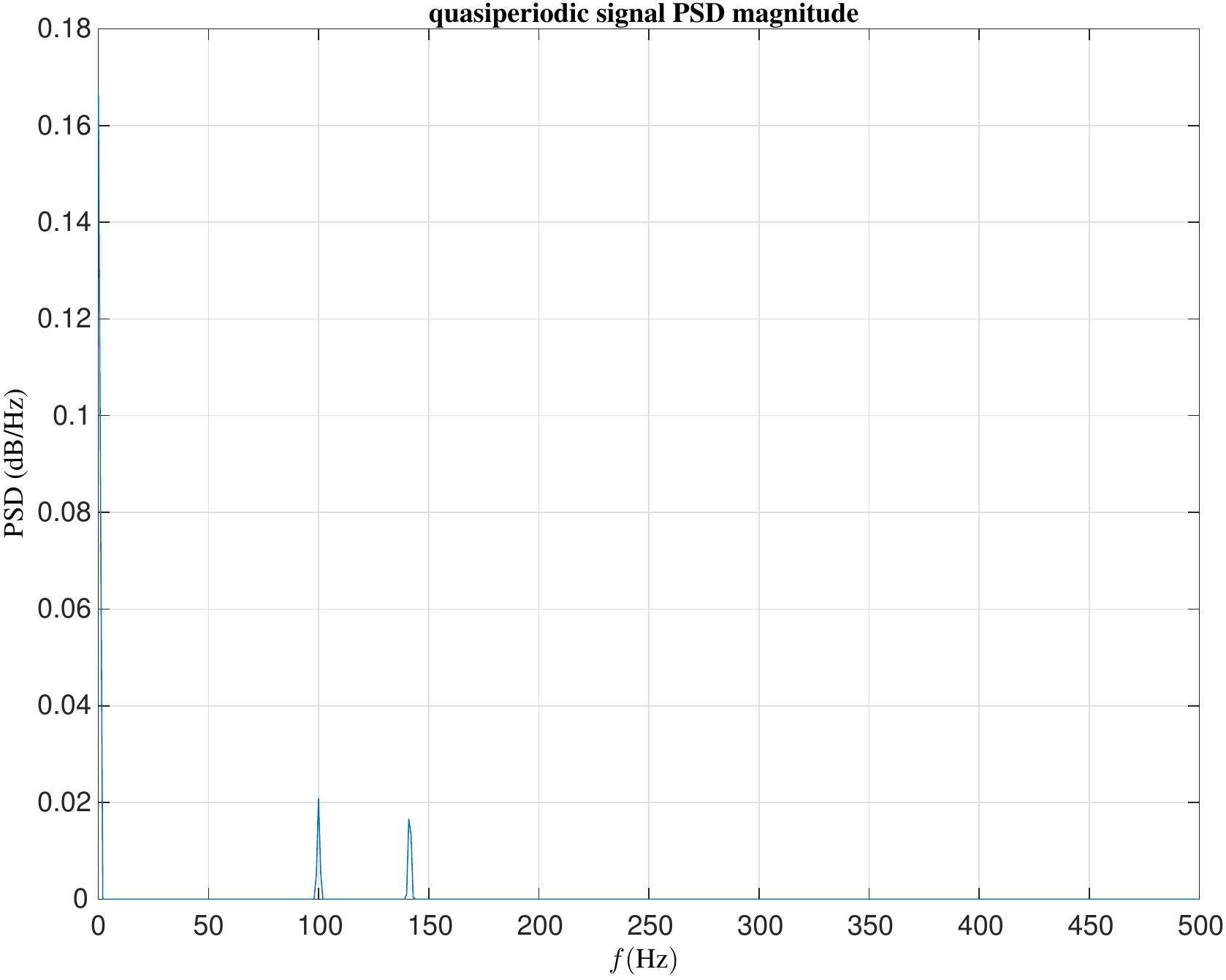}
\label{fig_second_case}}
\subfloat[]{\includegraphics[width=1.25in]{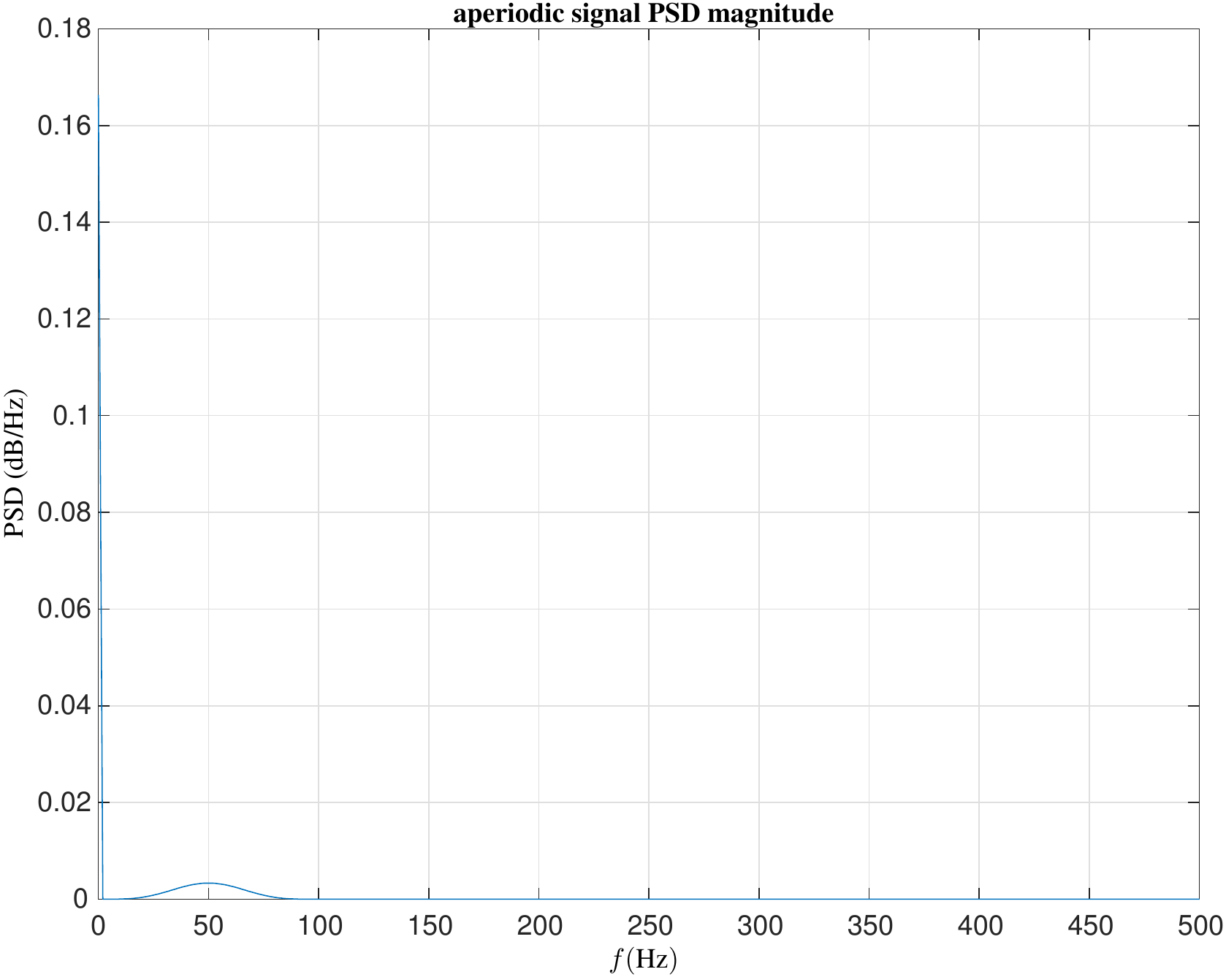}
\label{fig_third_case}}
\subfloat[]{\includegraphics[width=1.25in]{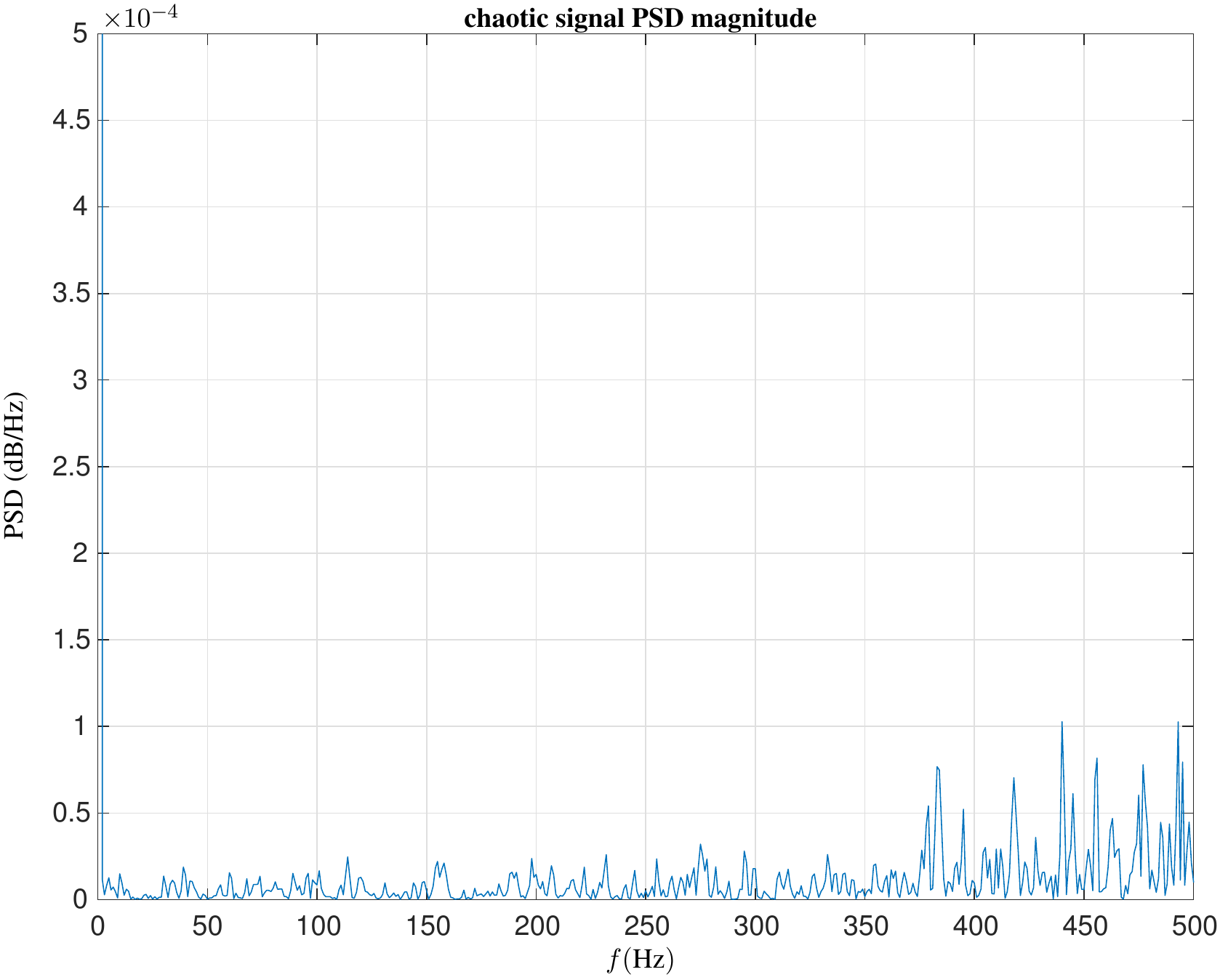}
\label{fig_four_case}}
\subfloat[]{\includegraphics[width=1.25in]{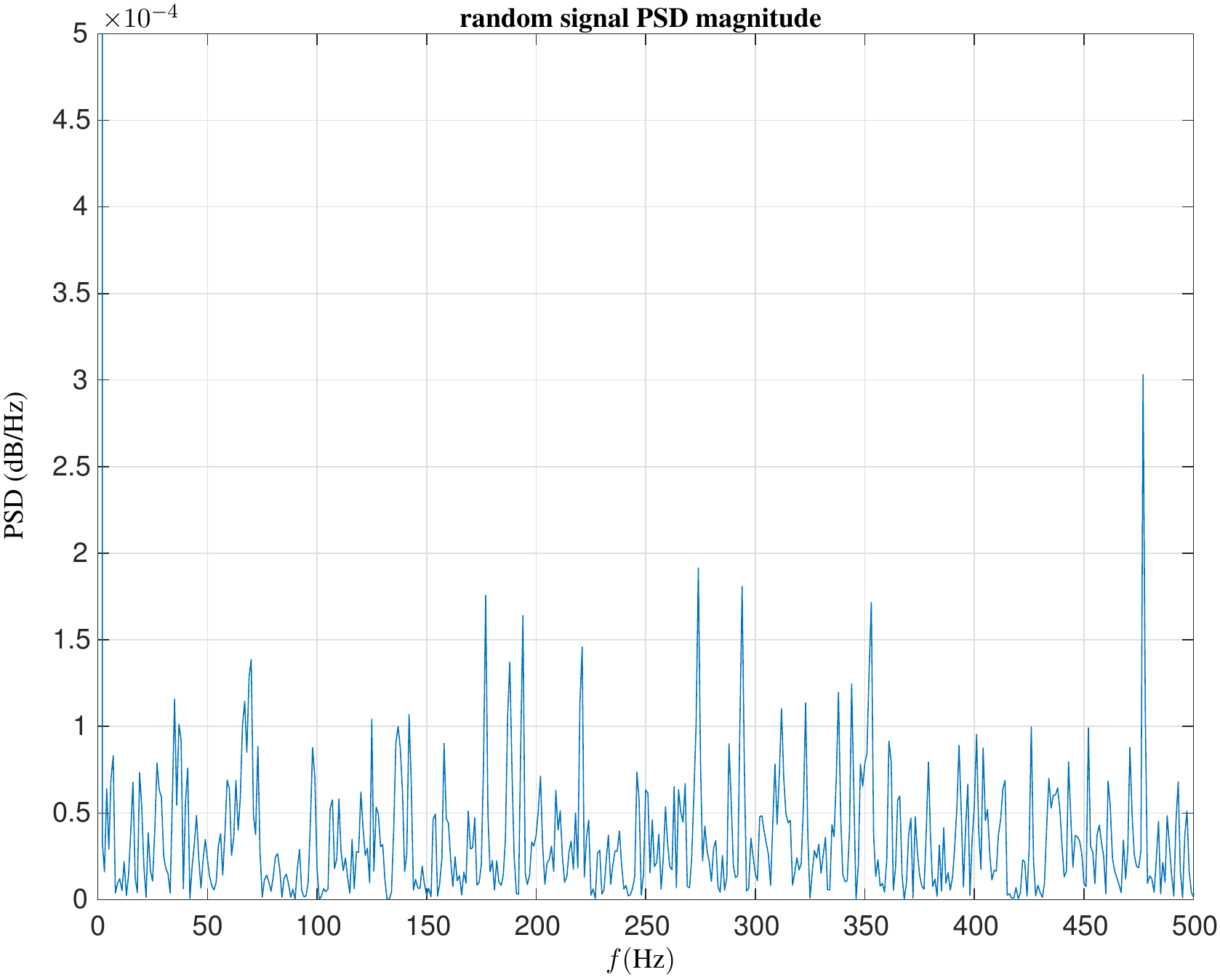}
\label{fig_five_case}}
\hfil
\subfloat[]{\includegraphics[width=1.25in]{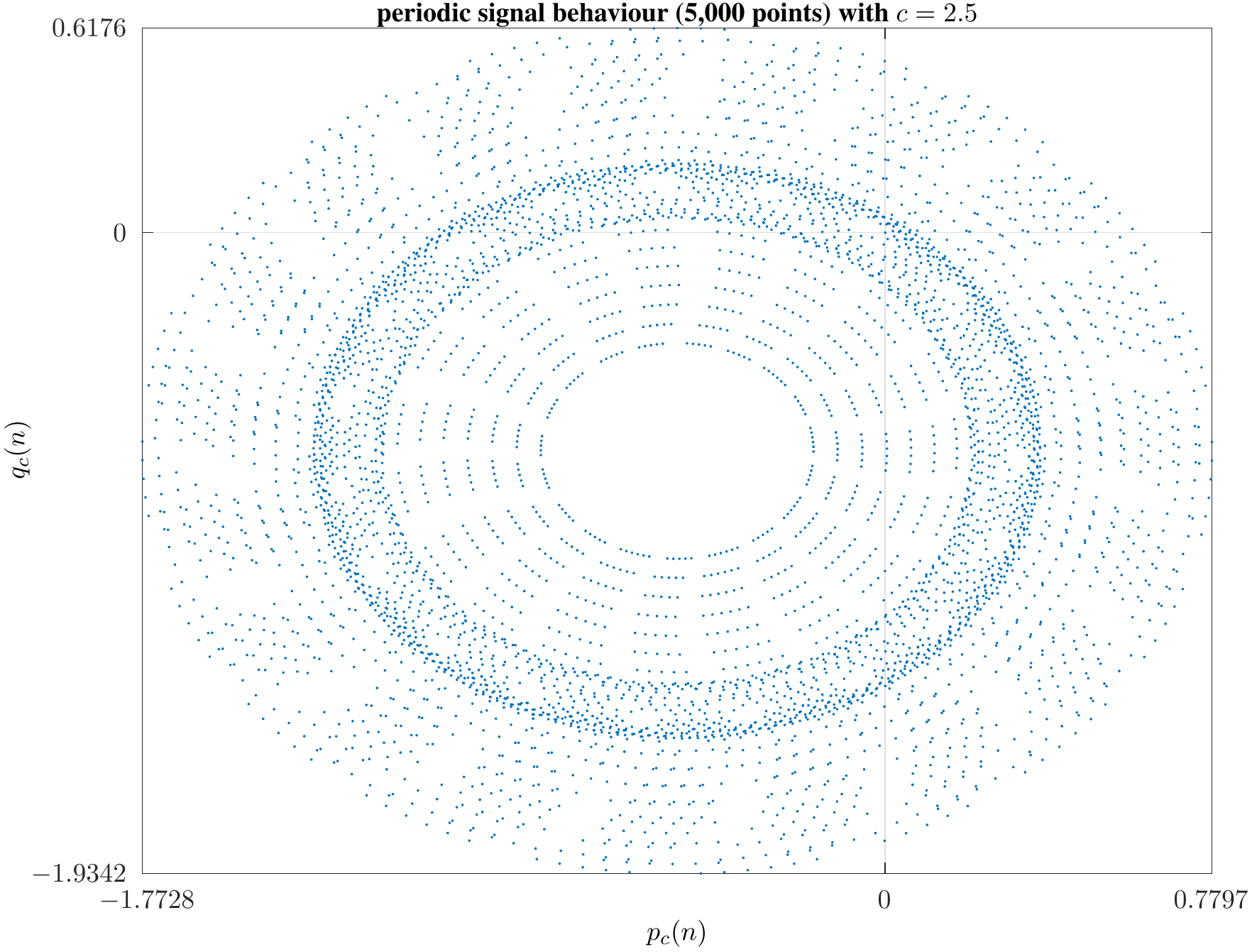}
\label{fig_six_case}}
\subfloat[]{\includegraphics[width=1.25in]{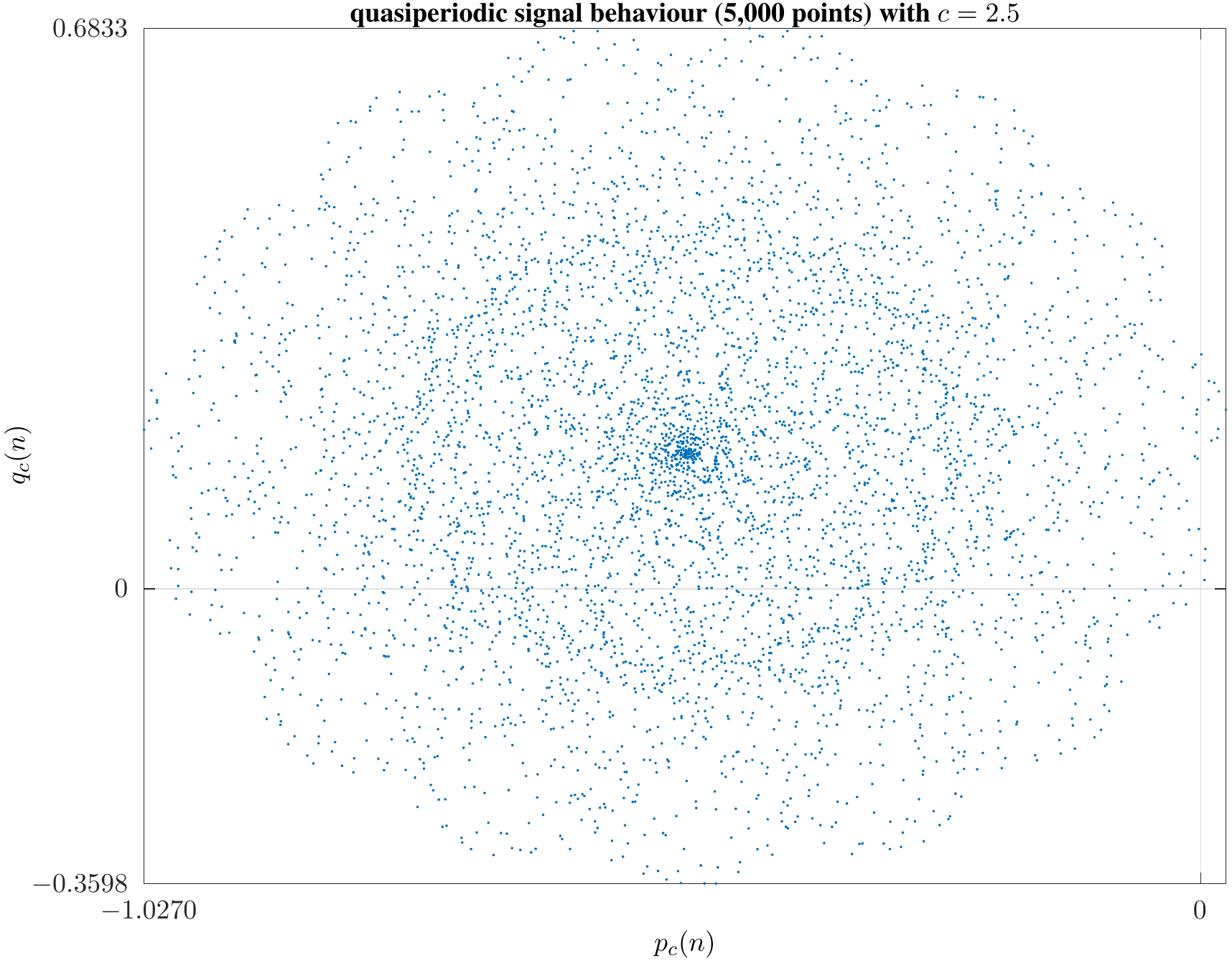}
\label{fig_seven_case}}
\subfloat[]{\includegraphics[width=1.25in]{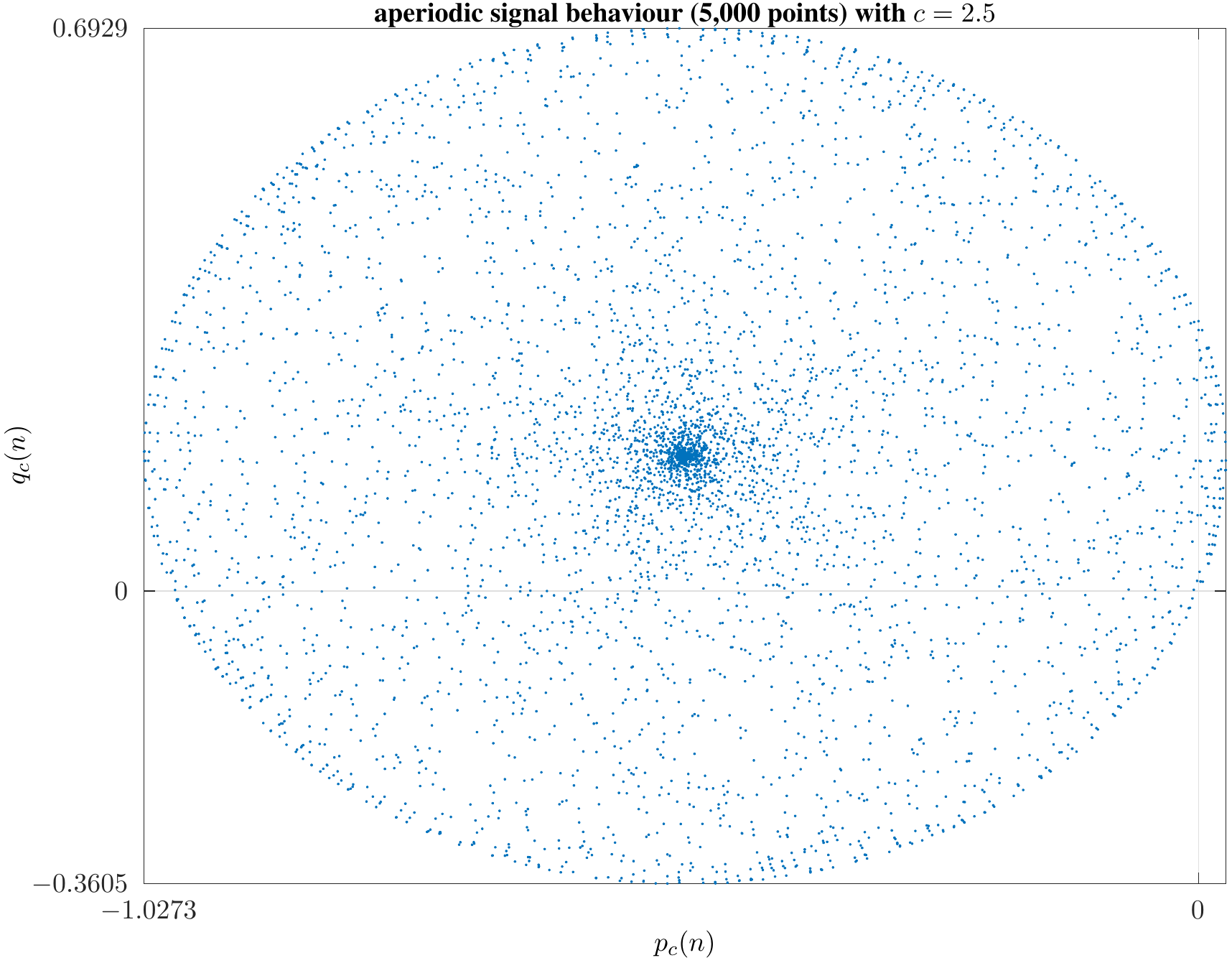}
\label{fig_eight_case}}
\subfloat[]{\includegraphics[width=1.25in]{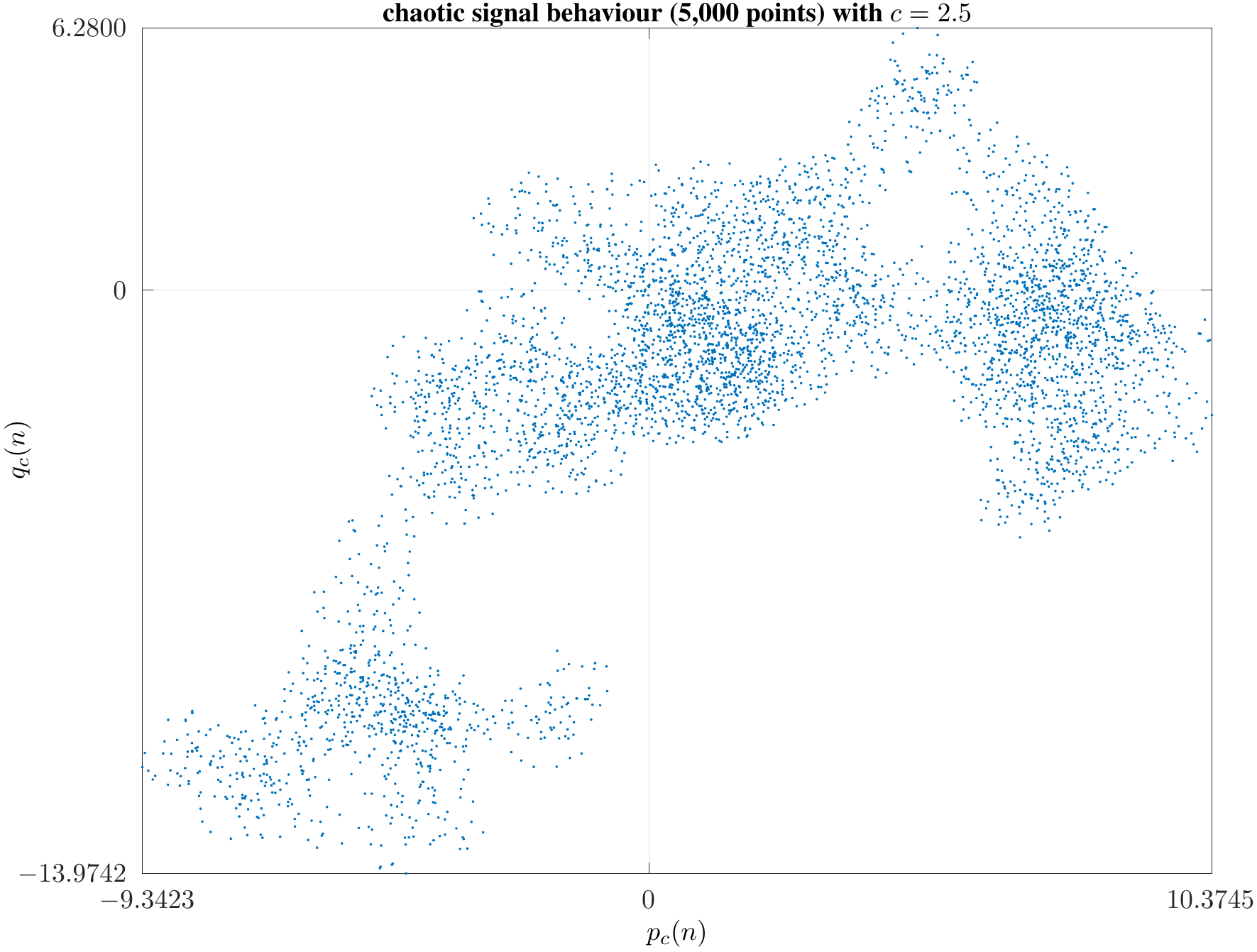}
\label{fig_nine_case}}
\subfloat[]{\includegraphics[width=1.25in]{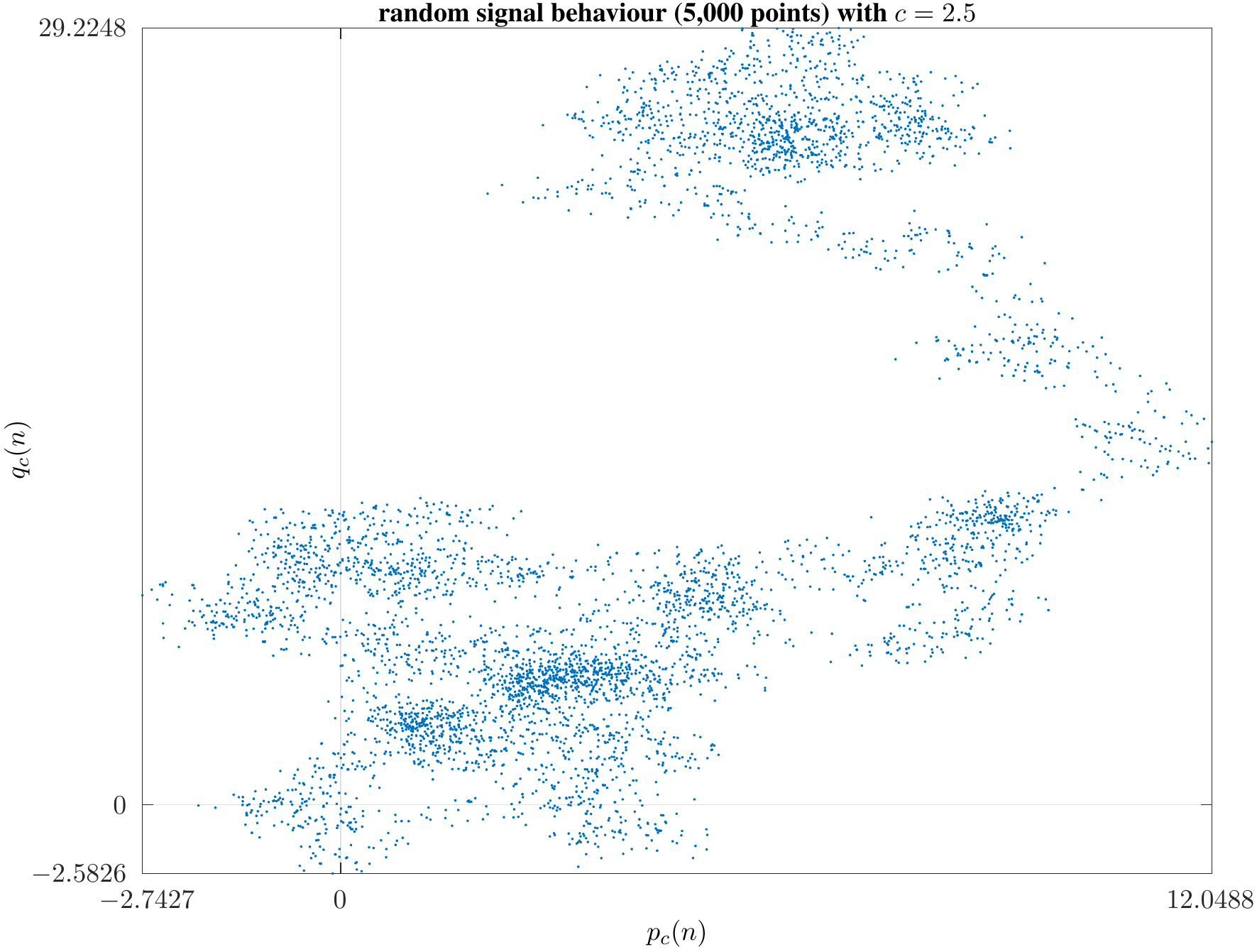}
\label{fig_ten_case}}
\hfil
\subfloat[]{\includegraphics[width=1.25in]{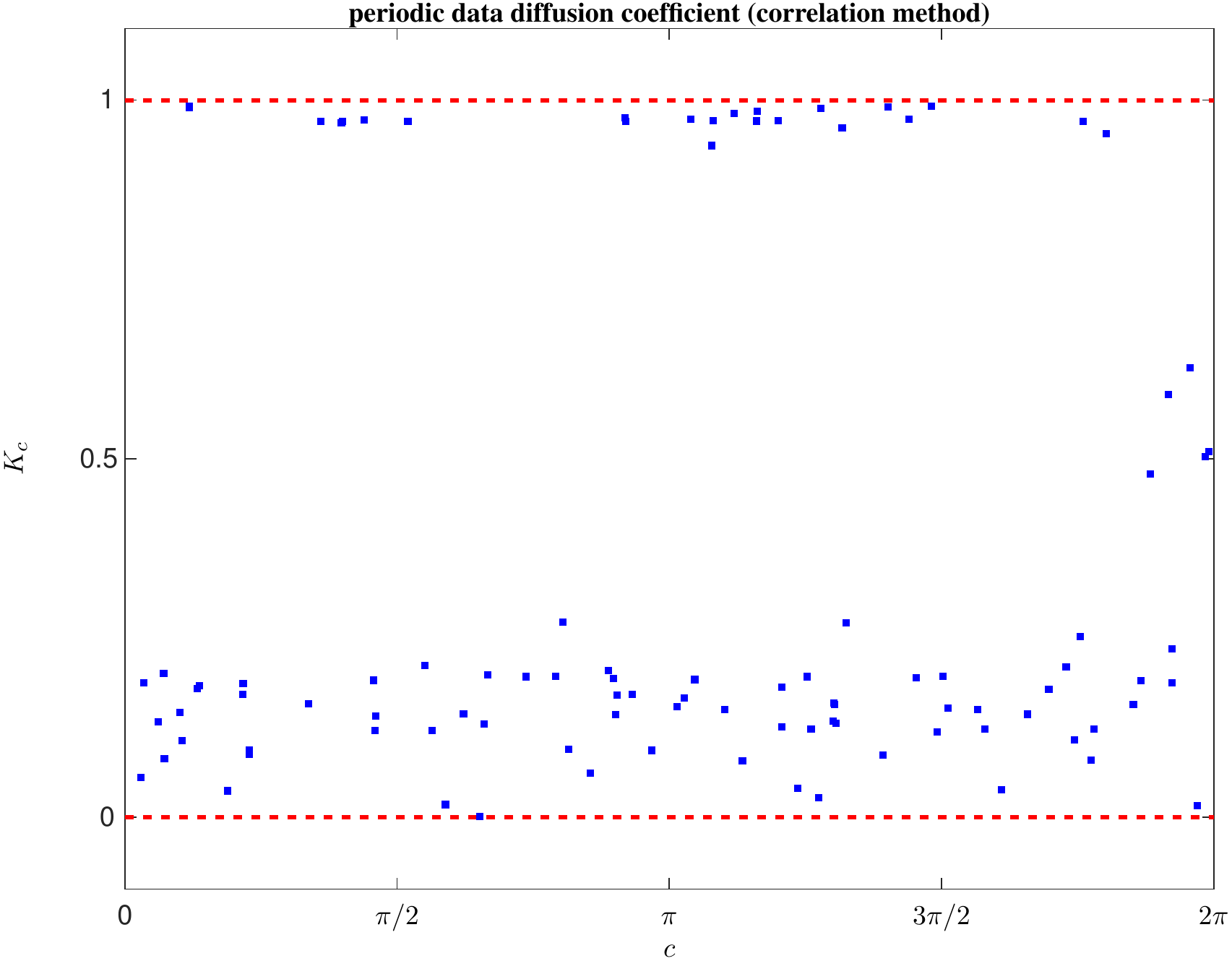}
\label{fig_eleven_case}}
\subfloat[]{\includegraphics[width=1.25in]{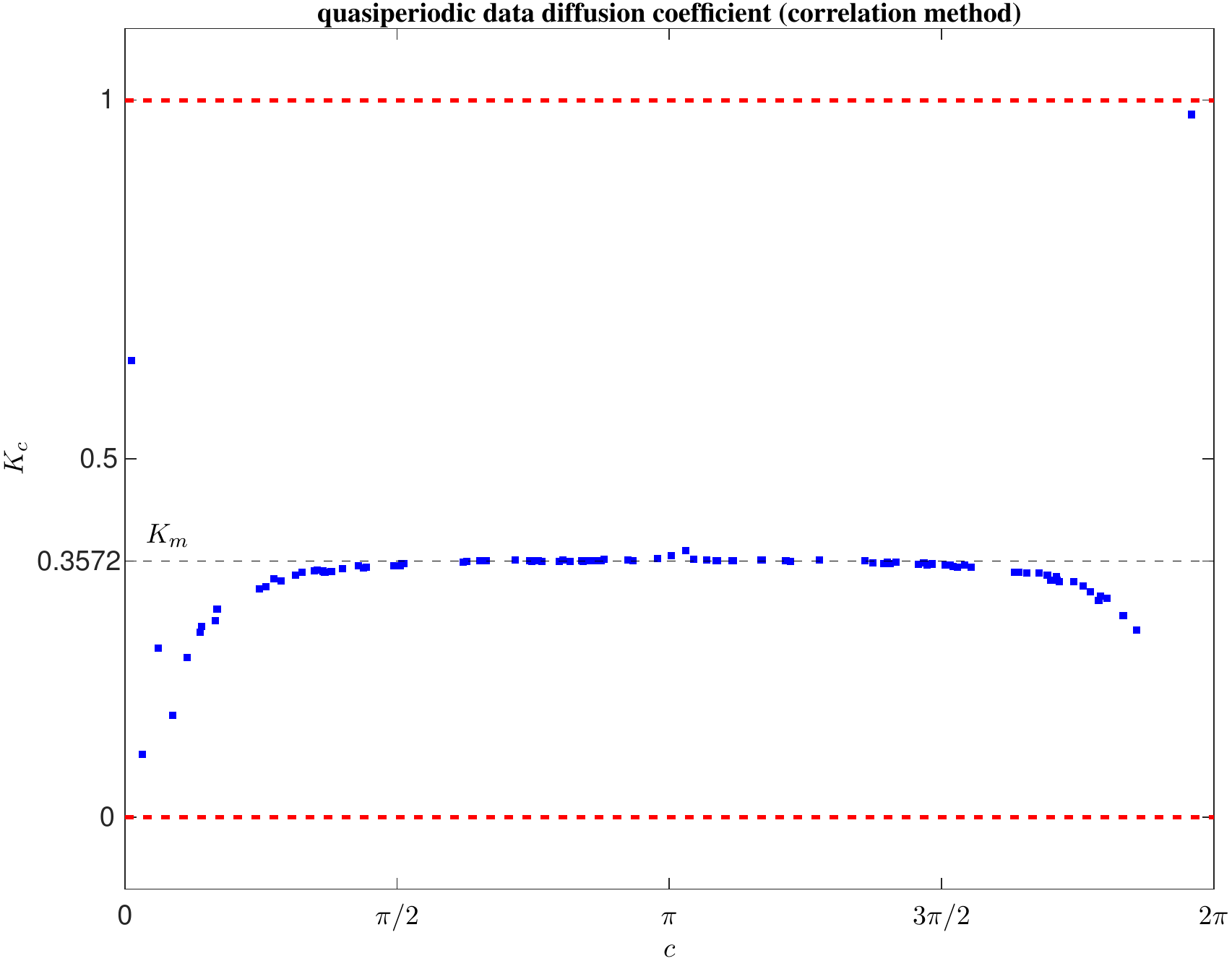}
\label{fig_twelve_case}}
\subfloat[]{\includegraphics[width=1.25in]{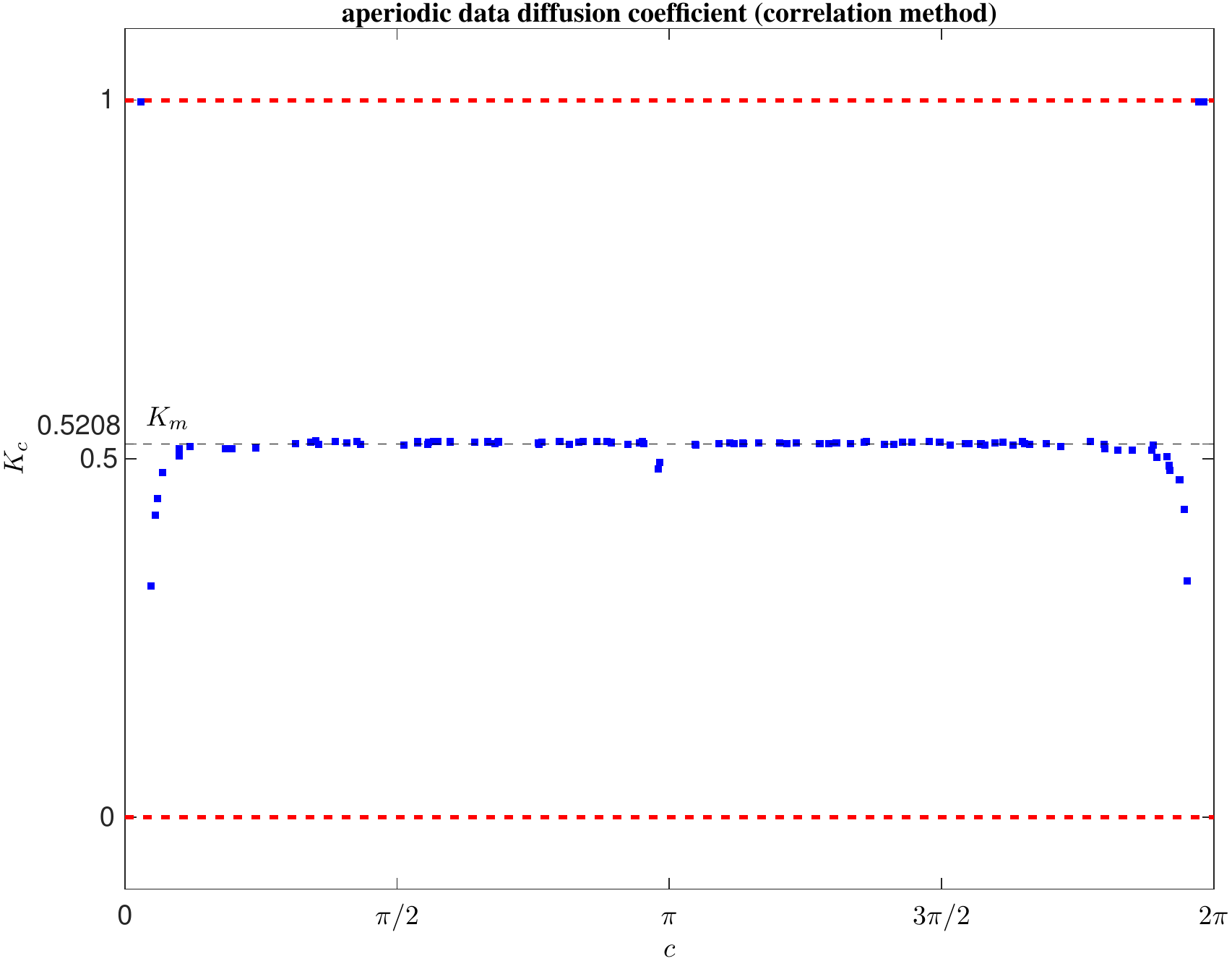}
\label{fig_thirteen_case}}
\subfloat[]{\includegraphics[width=1.25in]{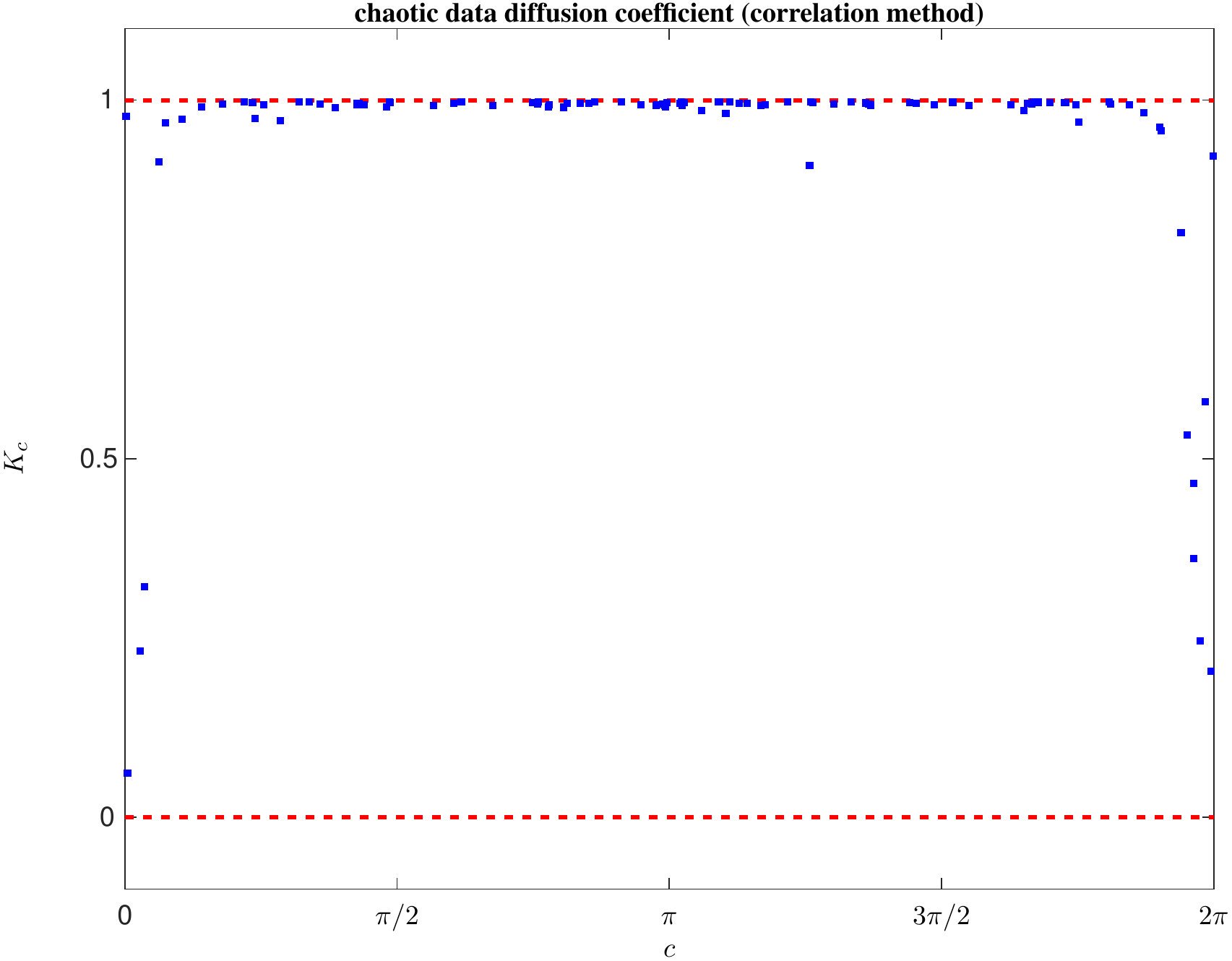}
\label{fig_fourteen_case}}
\subfloat[]{\includegraphics[width=1.25in]{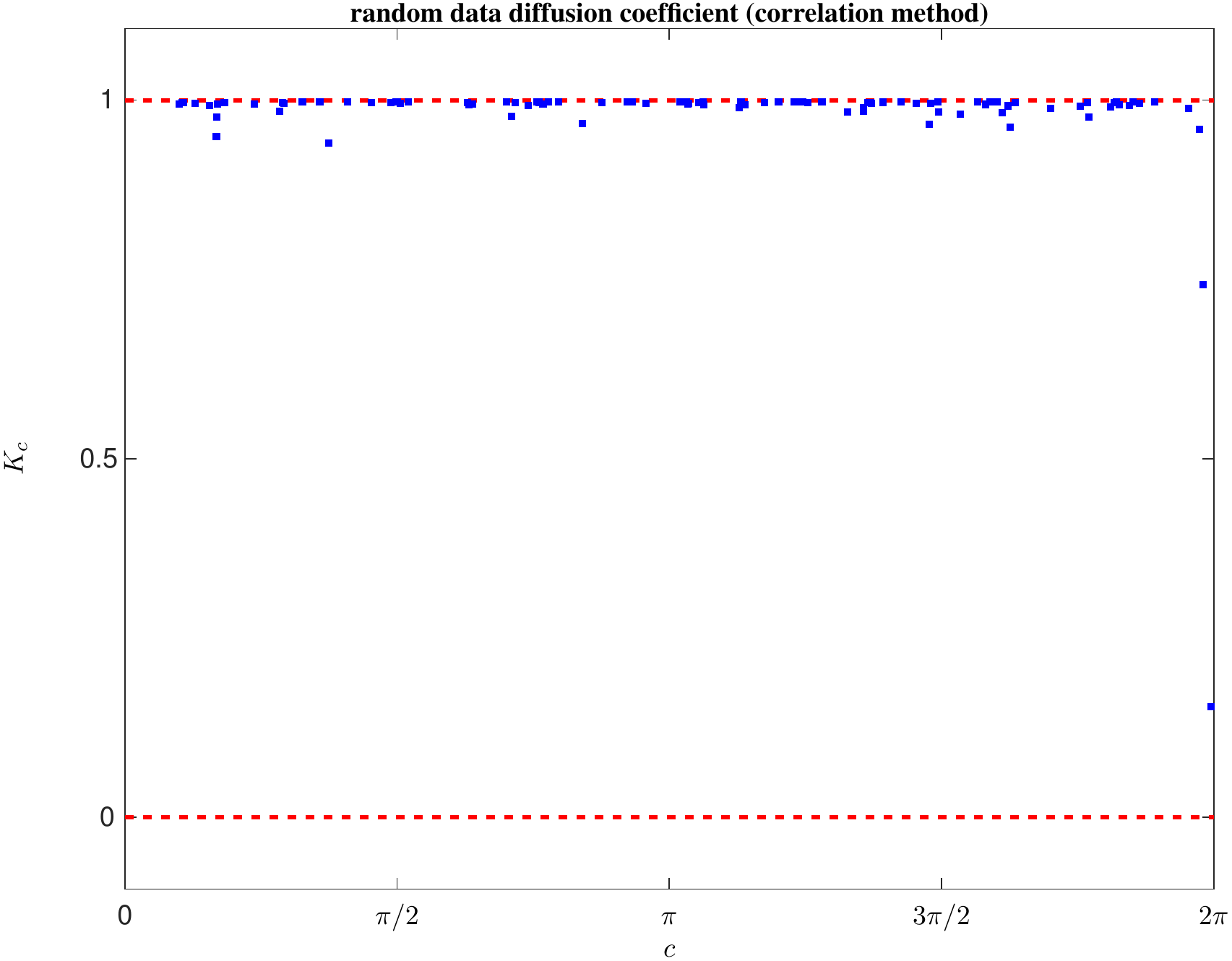}
\label{fig_fifteen_case}}
\hfil
\caption{Each column corresponds with a reference signal: periodic, quasi-periodic, aperiodic, chaotic, and random. (a)-(e) Power Spectrum Density (PSD); (f)-(j) Representation of equation \eqref{eqn:SD01} for $c=2.5$; (k)-(o) $K_{c}$ results of  0--1 test, for 100 random $c$ values, with correlation method.} 
\label{fig:F01}
\end{figure*}

\section{Results and discussions}\label{sec:RESULTS}
In Fig. \ref{fig:F01} are shown the most representative signals with the results obtained when the 0--1 test is applied as indicated in Table \ref{tab:T01}. As here we have not too much space, we decide that the power spectrum density (PSD) gives enough information to compare the signals. Their representation is in Fig. \ref{fig:F01}(a)-(e), as we can see there is a big difference between the most regular, the periodic one, and the chaotic one; and not that much difference between the deterministic chaos and the random behavior.  Also is relevant to compare the periodic and quasi-periodic signals where both have two frequencies, but with a rational and irrational relation, respectively; the following graphs, in Fig. \ref{fig:F01}, show the consequence of it.
\begin{figure*}[!t]
\centering
\subfloat[]{\includegraphics[width=1.25in]{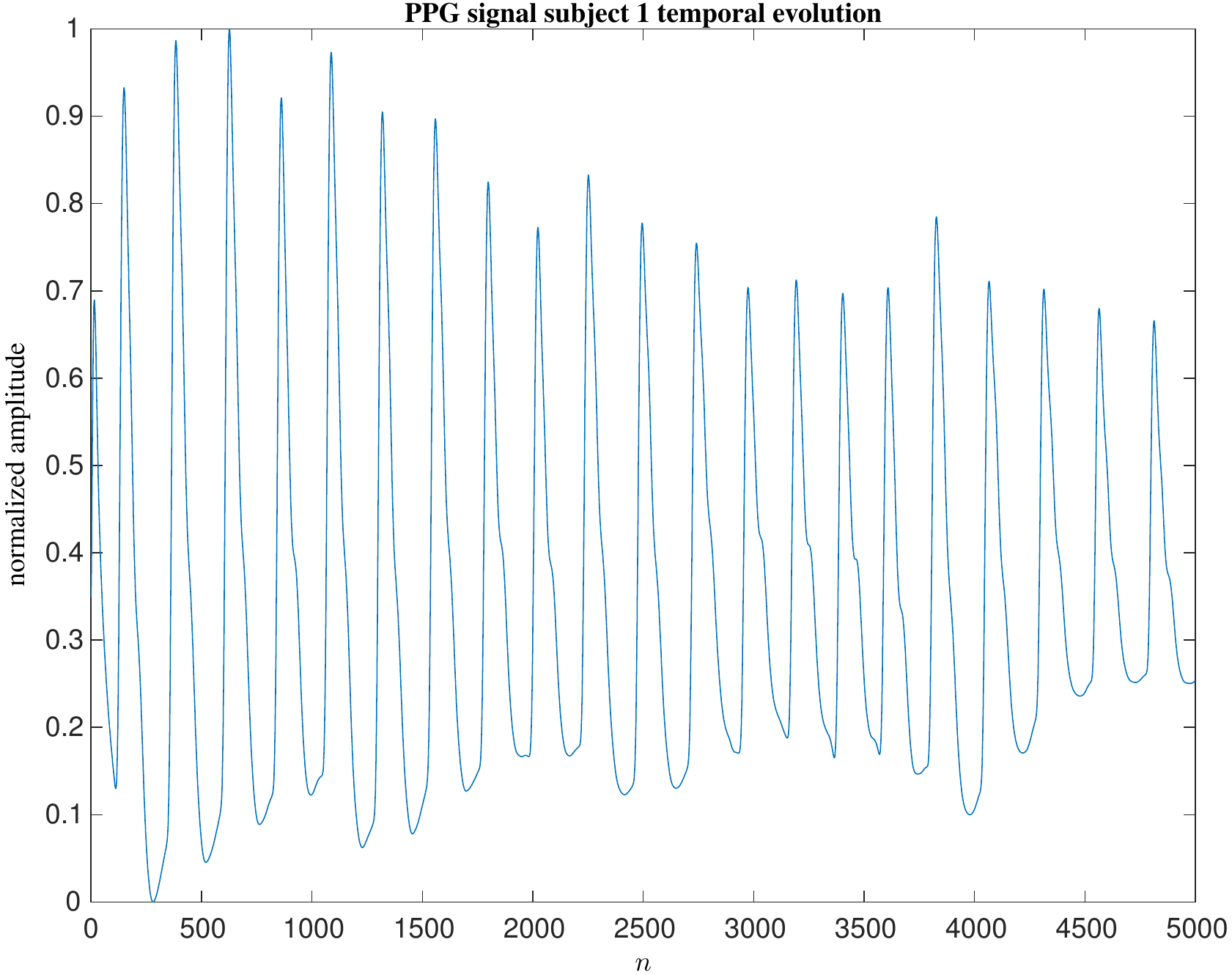}
\label{fig_first_case}}
\subfloat[]{\includegraphics[width=1.25in]{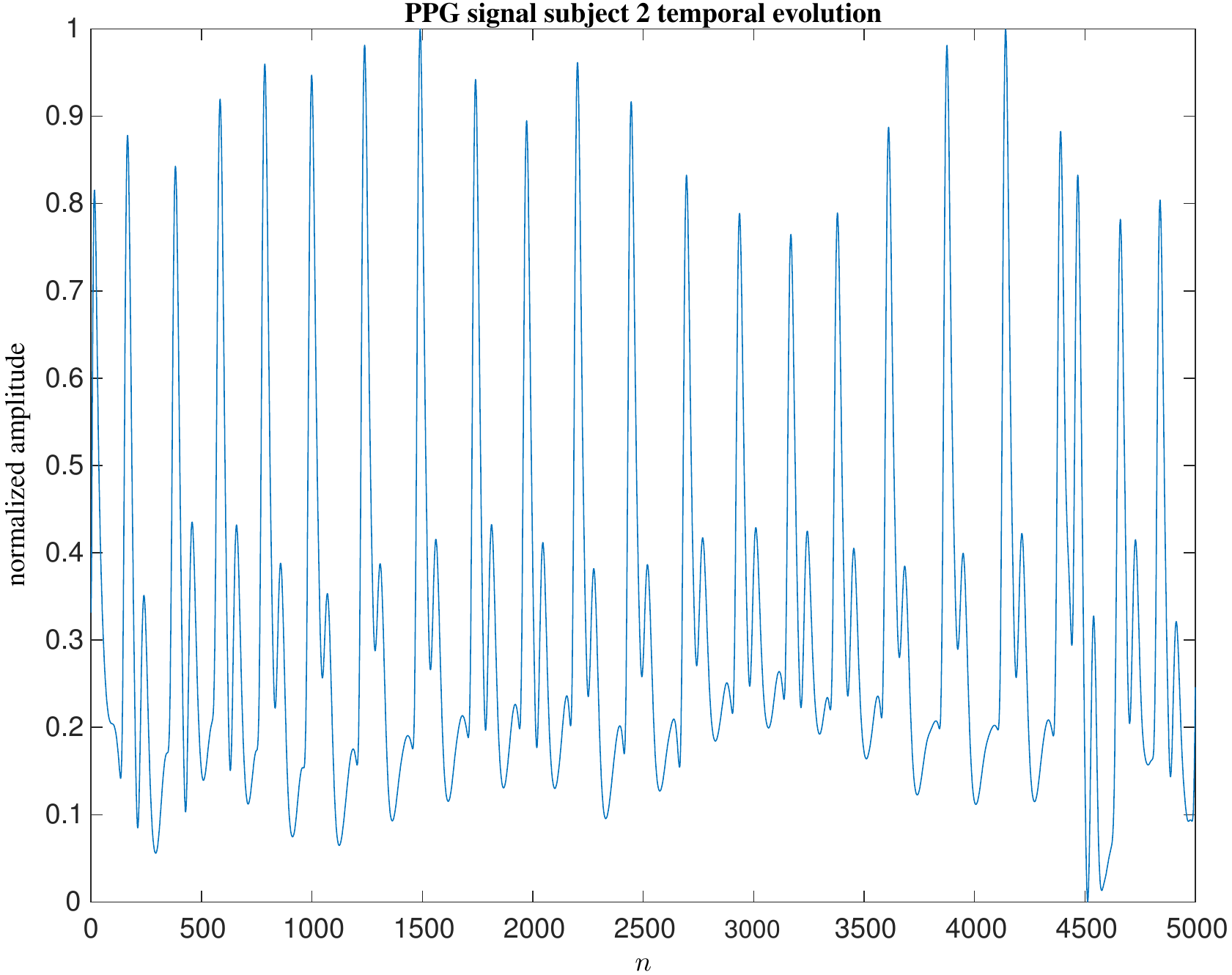}
\label{fig_second_case}}
\subfloat[]{\includegraphics[width=1.25in]{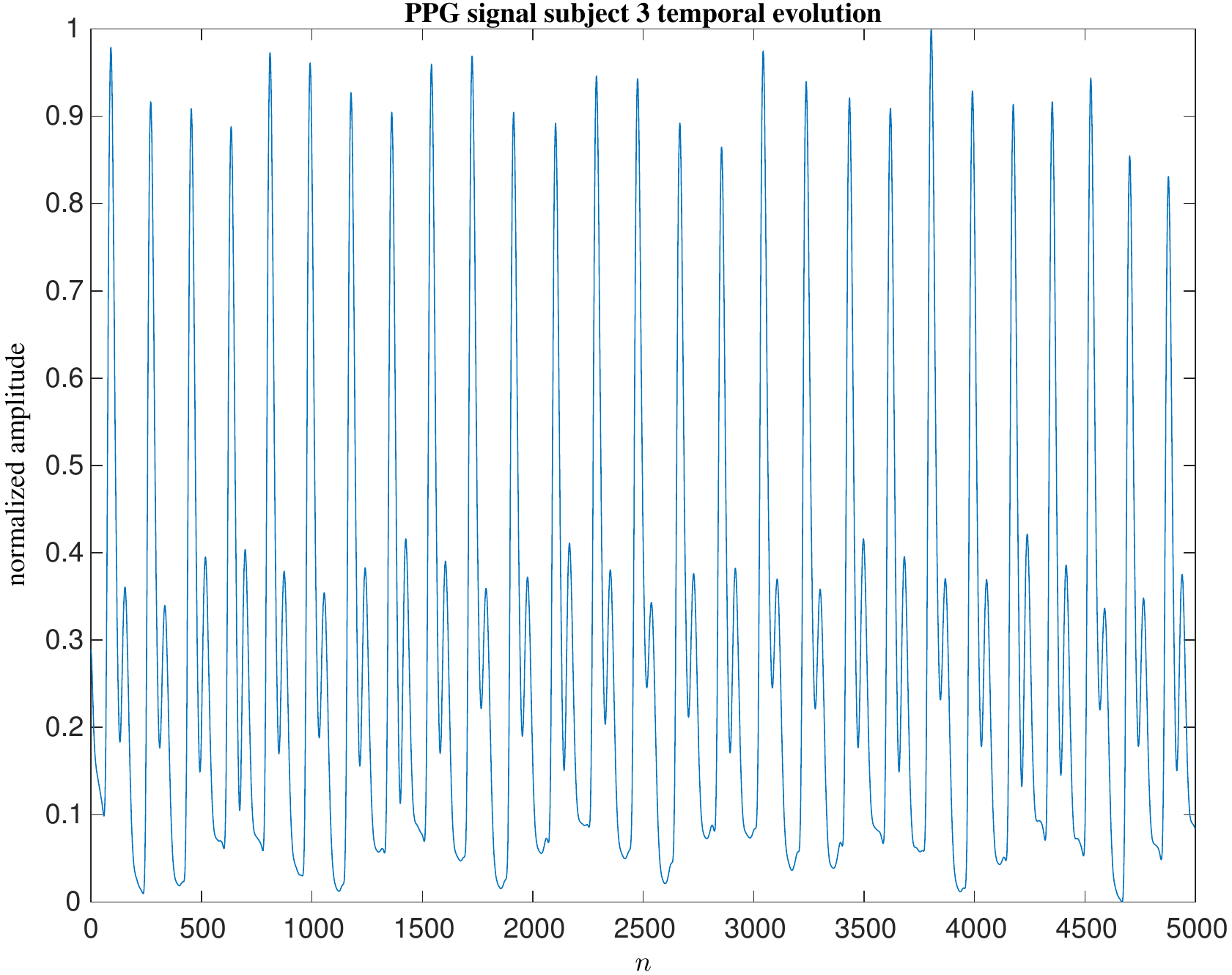}
\label{fig_third_case}}
\subfloat[]{\includegraphics[width=1.25in]{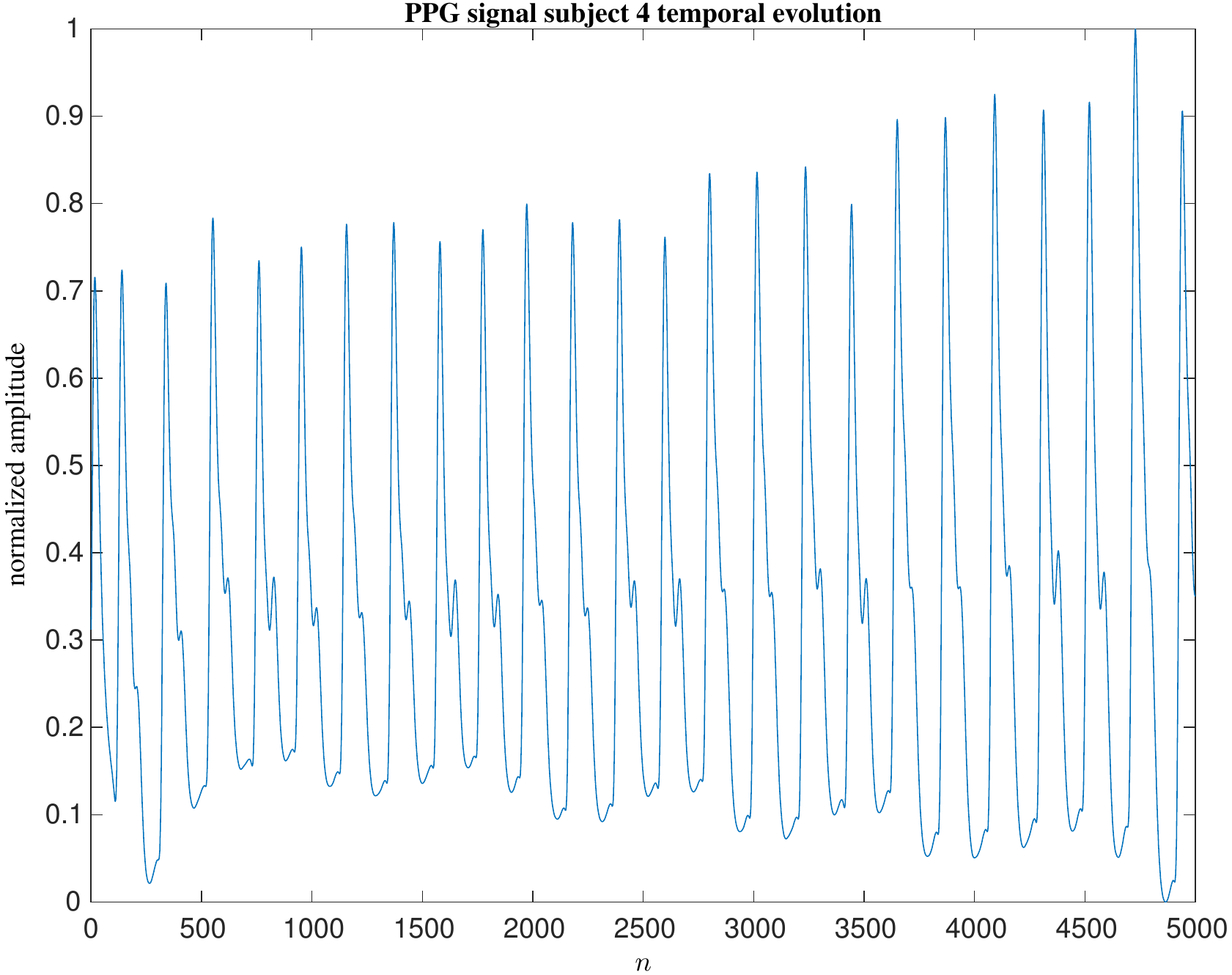}
\label{fig_four_case}}
\subfloat[]{\includegraphics[width=1.25in]{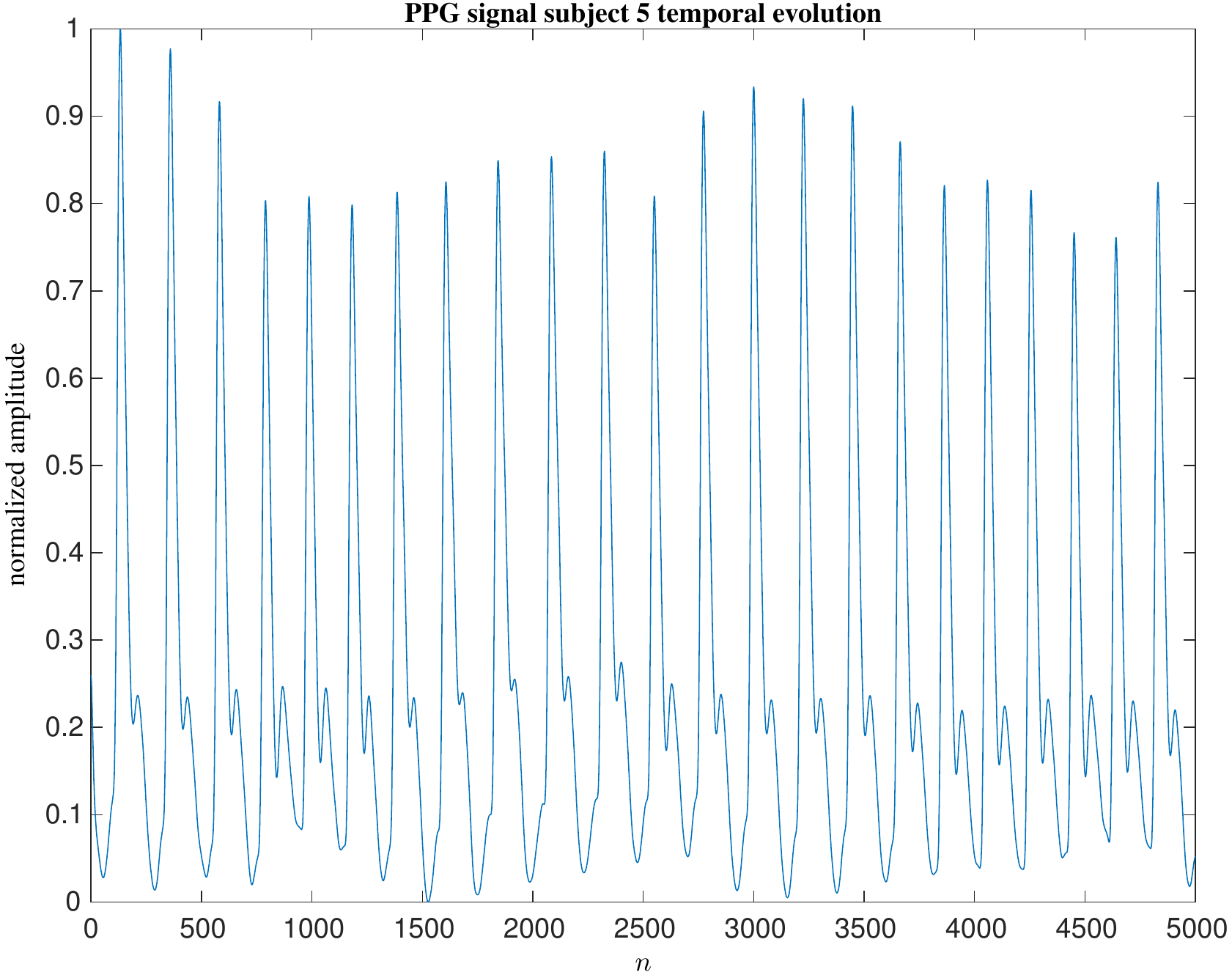}
\label{fig_five_case}}
\hfil
\subfloat[]{\includegraphics[width=1.25in]{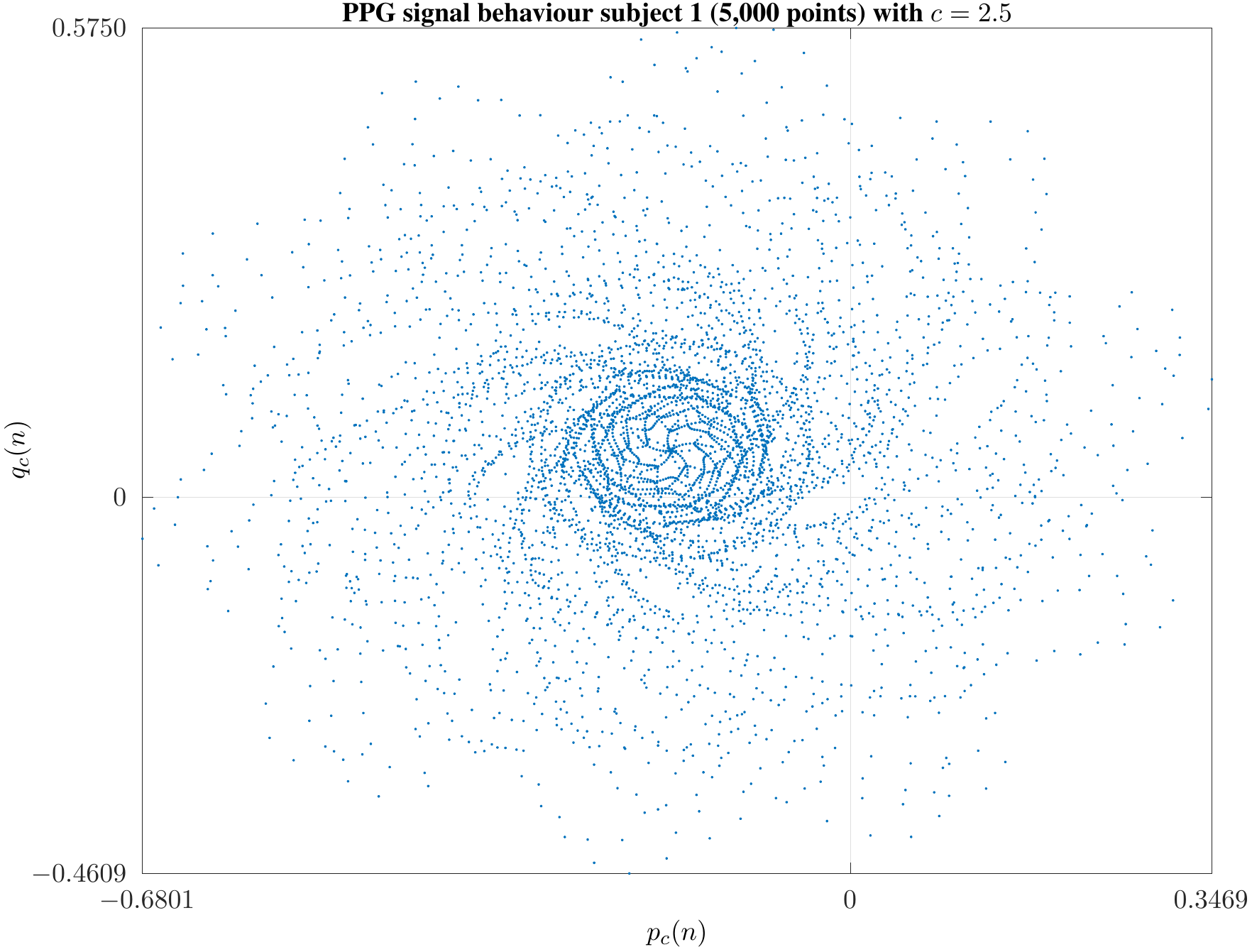}
\label{fig_six_case}}
\subfloat[]{\includegraphics[width=1.25in]{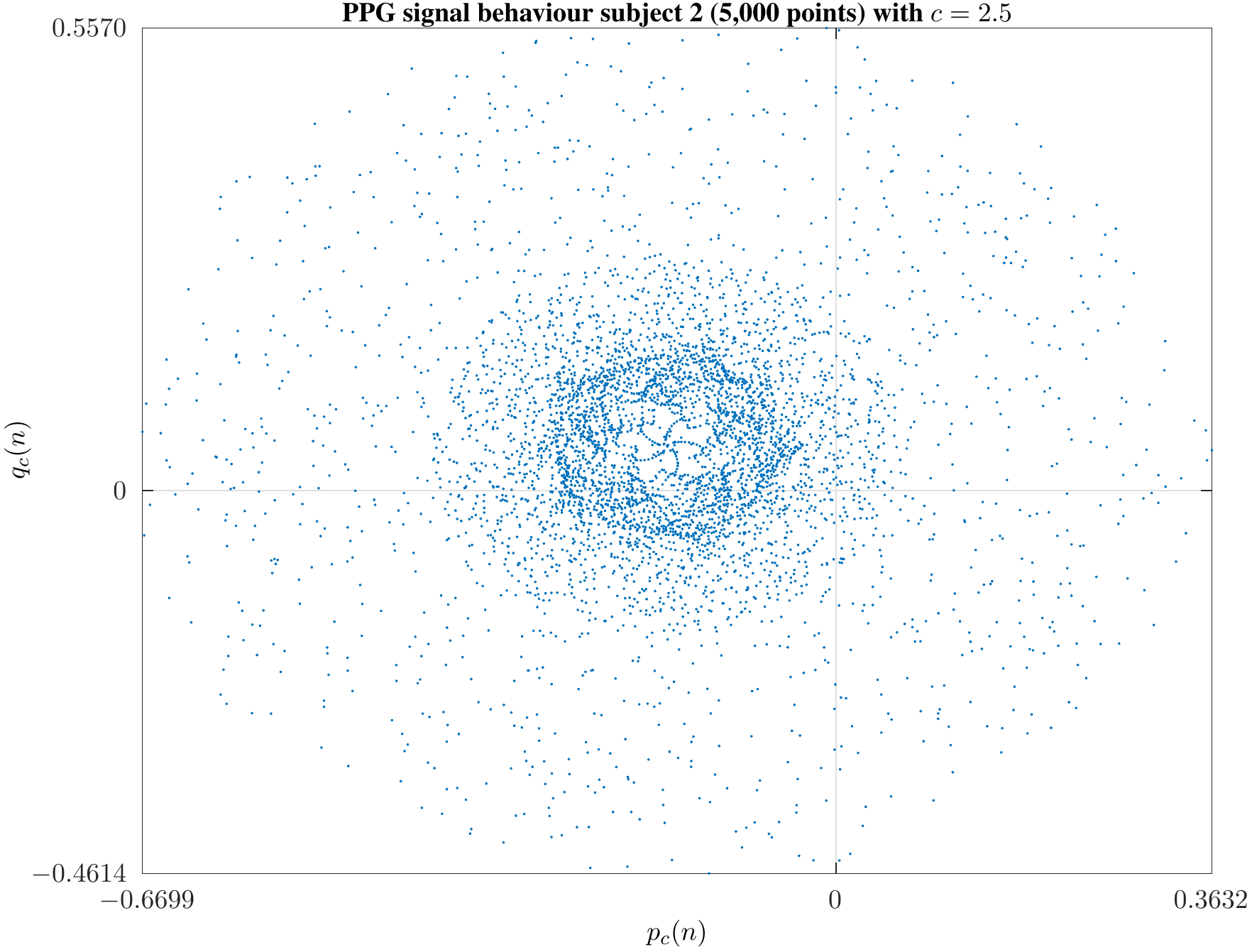}
\label{fig_seven_case}}
\subfloat[]{\includegraphics[width=1.25in]{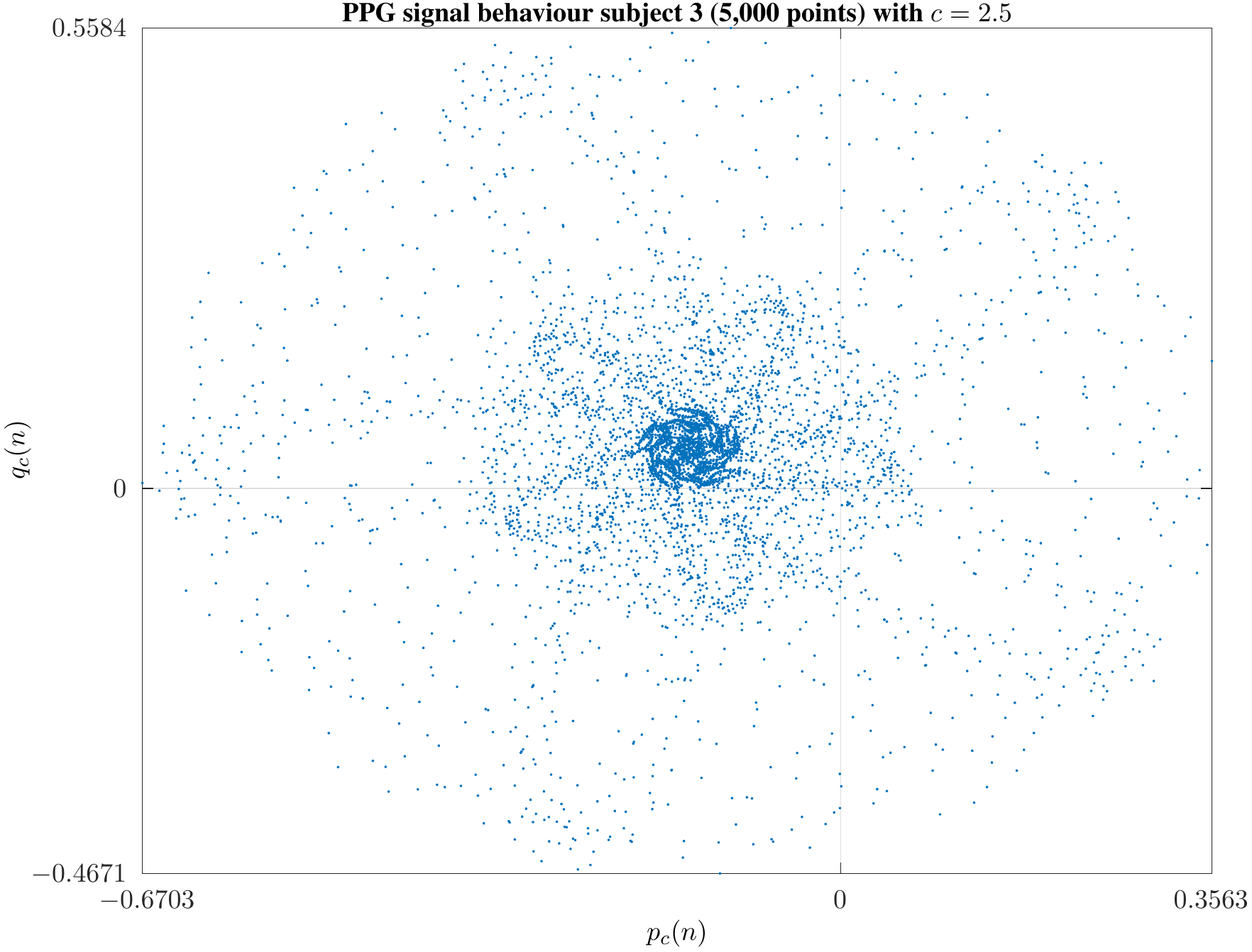}
\label{fig_eight_case}}
\subfloat[]{\includegraphics[width=1.25in]{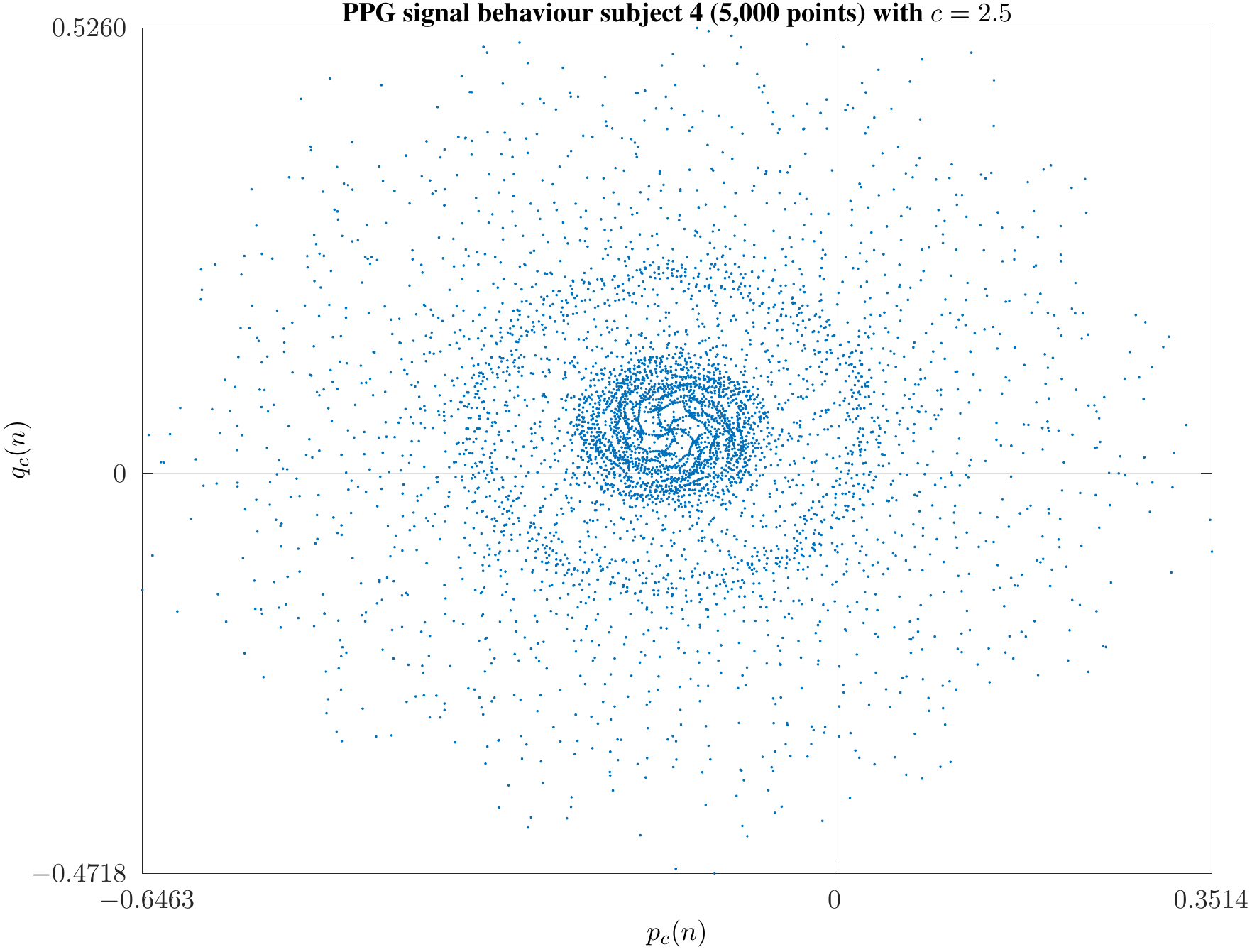}
\label{fig_nine_case}}
\subfloat[]{\includegraphics[width=1.25in]{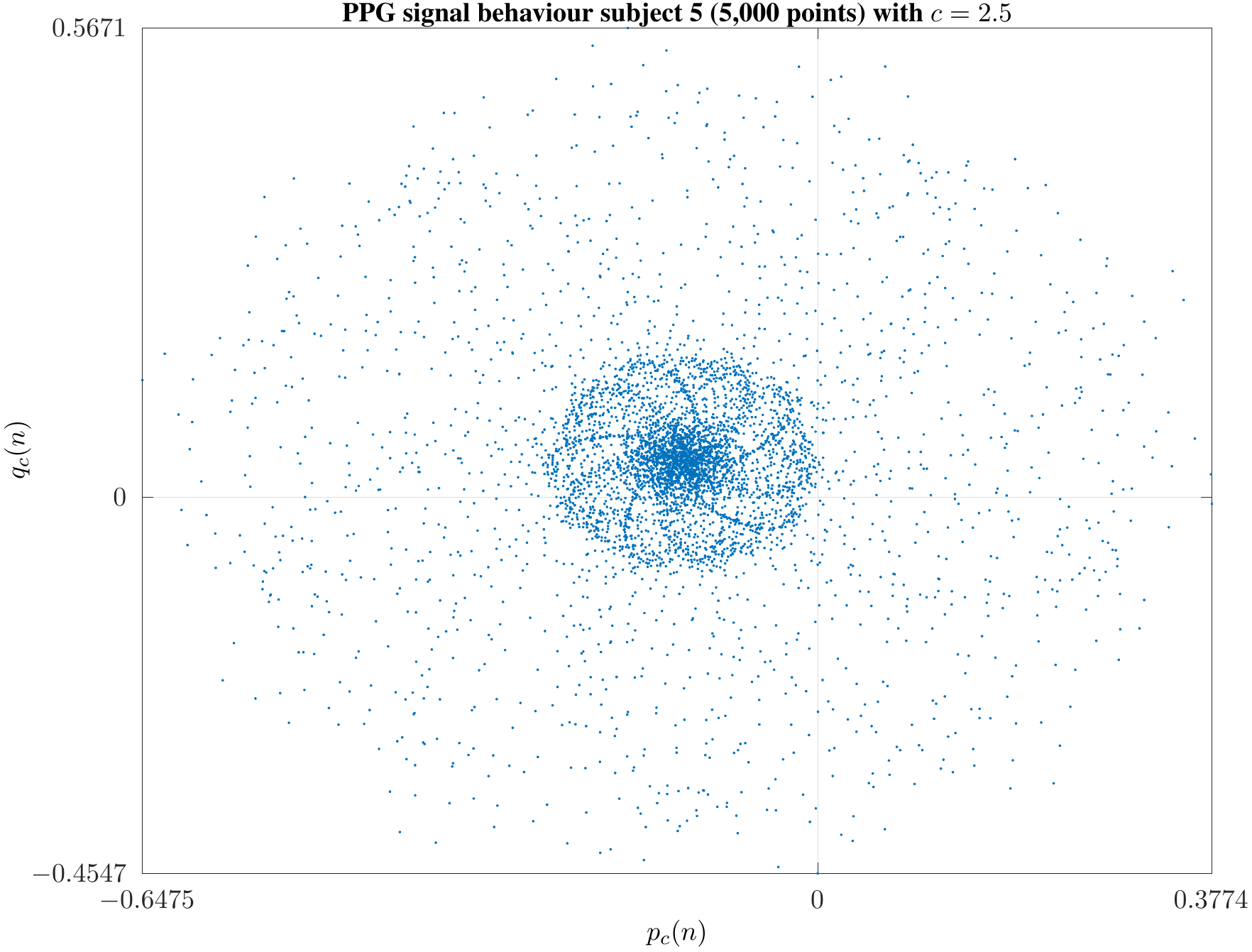}
\label{fig_ten_case}}
\hfil
\subfloat[]{\includegraphics[width=1.25in]{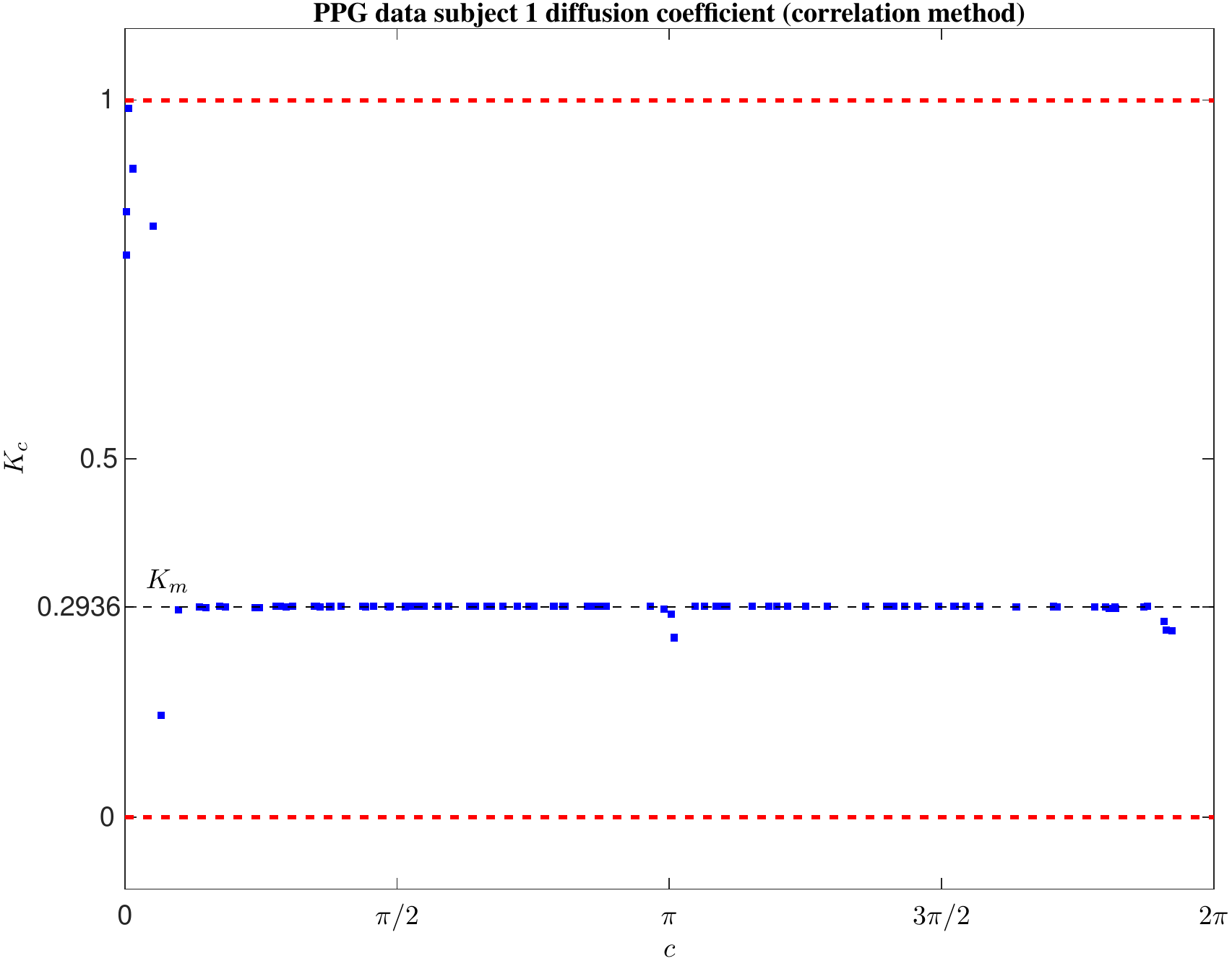}
\label{fig_six_case}}
\subfloat[]{\includegraphics[width=1.25in]{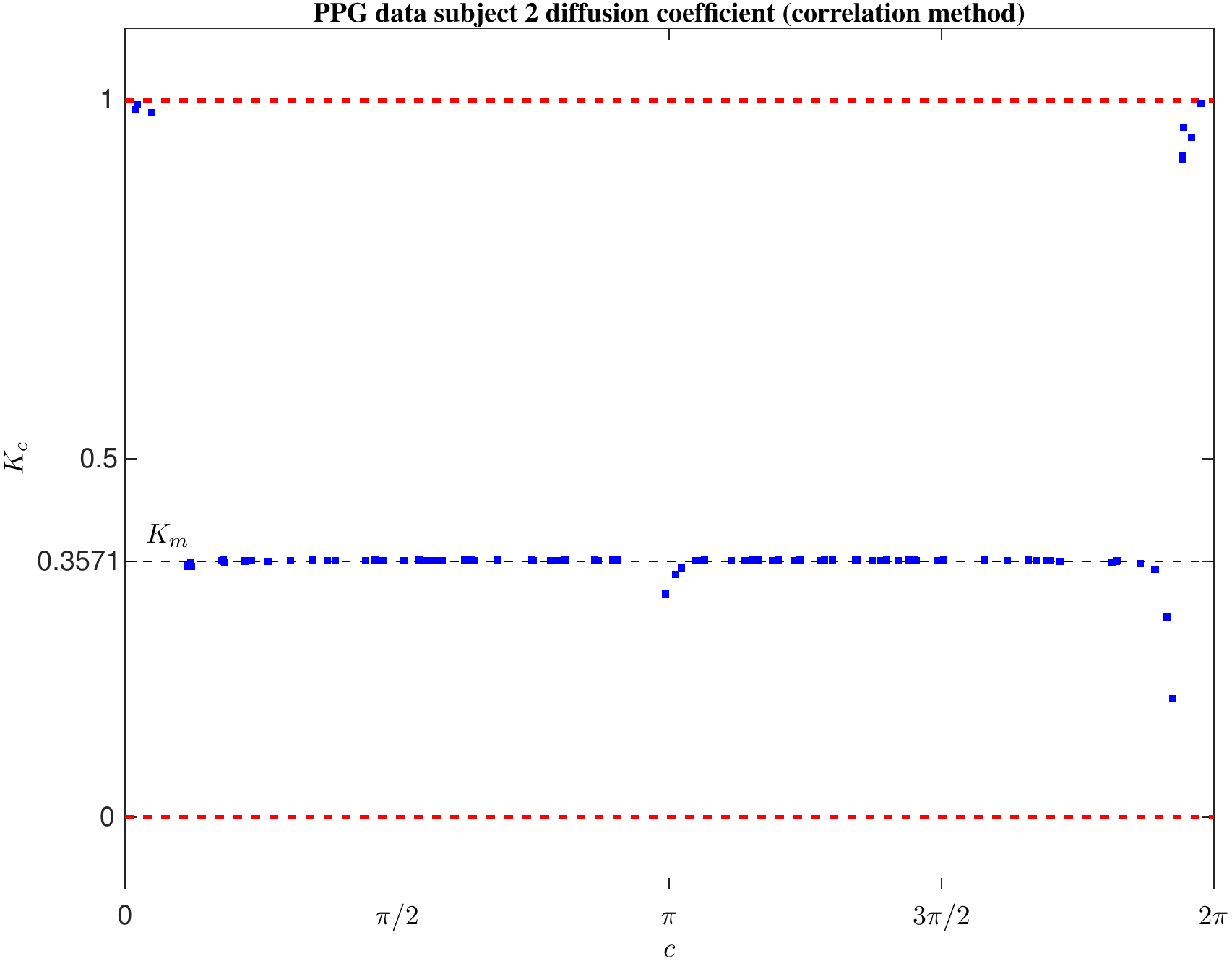}
\label{fig_seven_case}}
\subfloat[]{\includegraphics[width=1.25in]{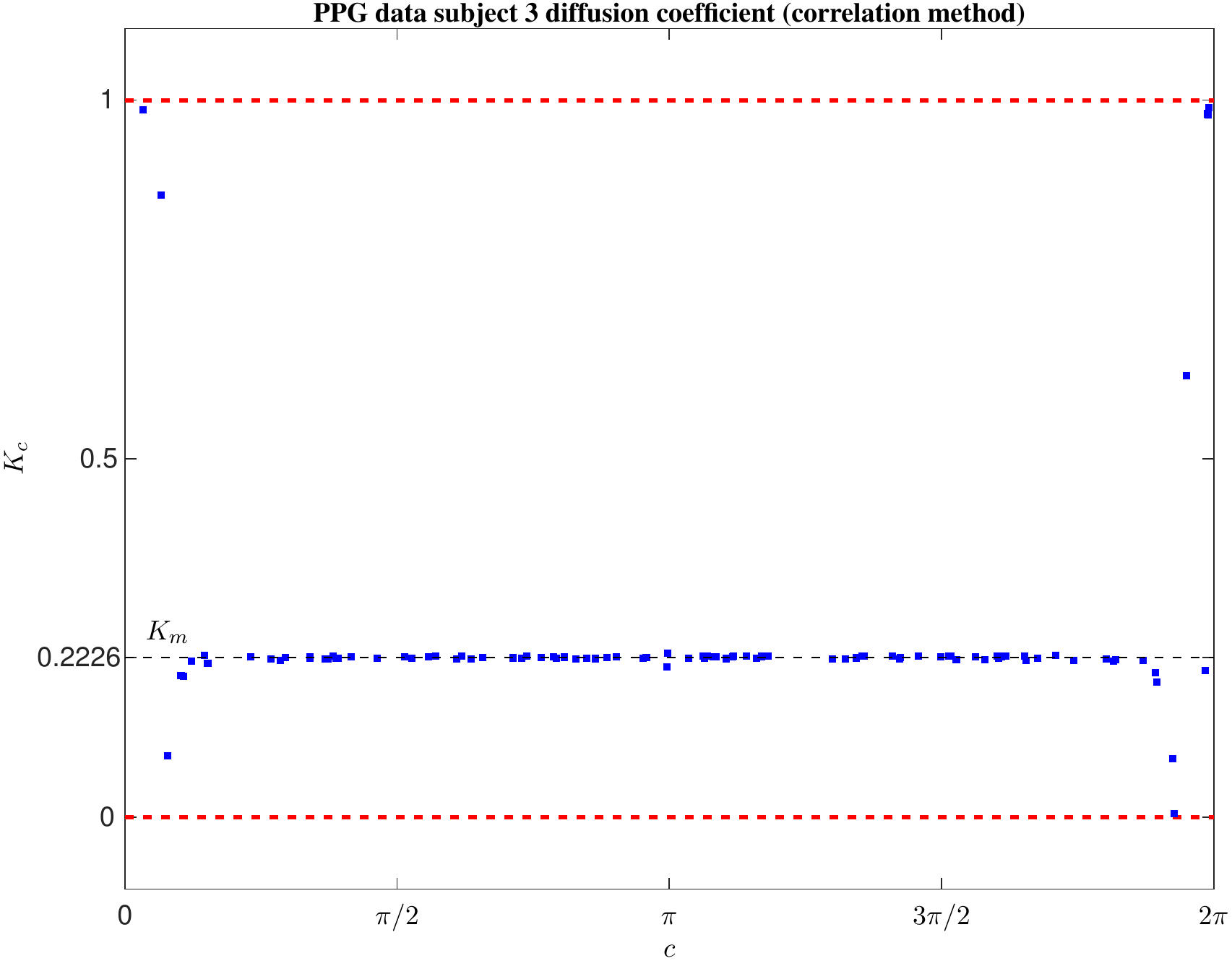}
\label{fig_eight_case}}
\subfloat[]{\includegraphics[width=1.25in]{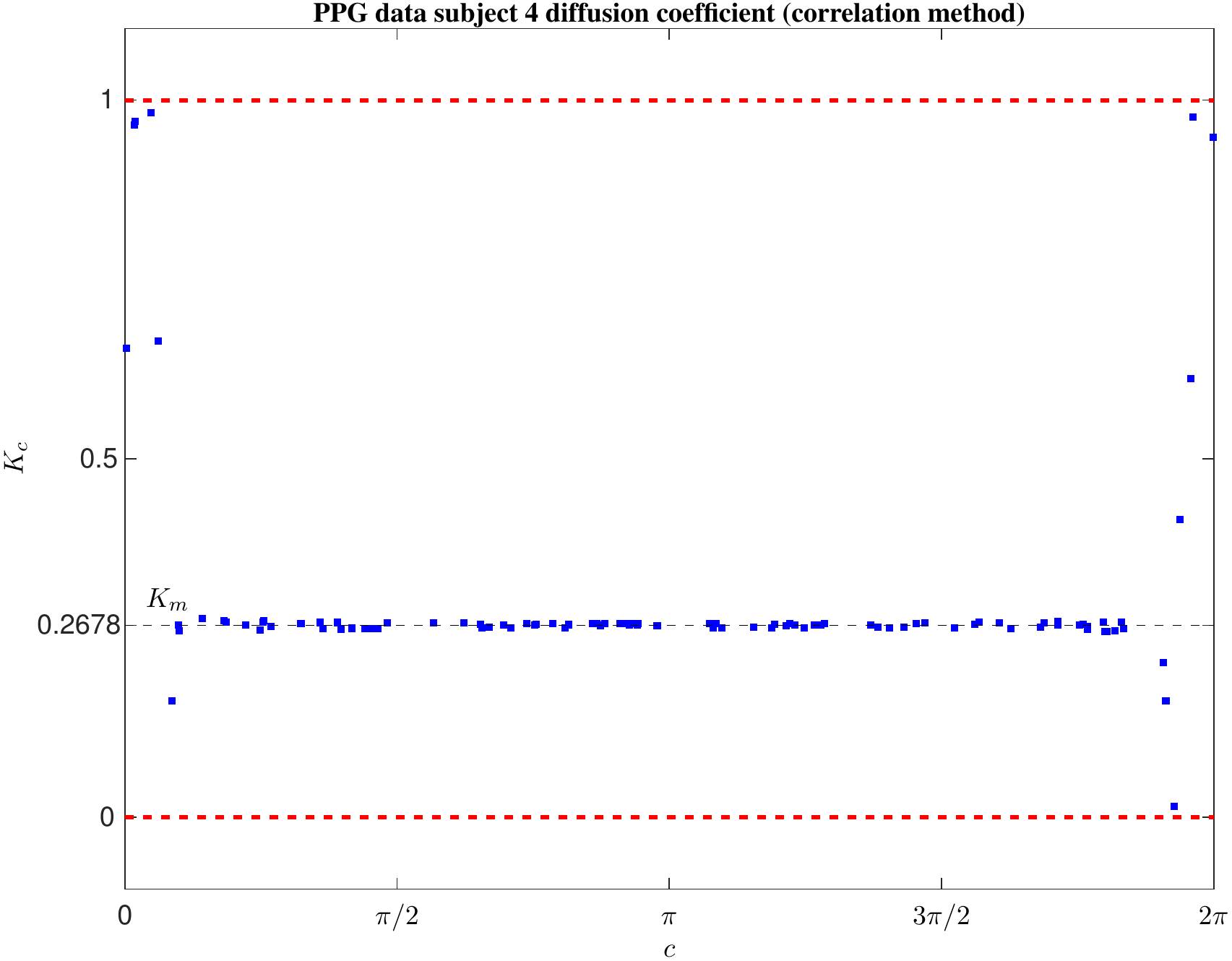}
\label{fig_nine_case}}
\subfloat[]{\includegraphics[width=1.25in]{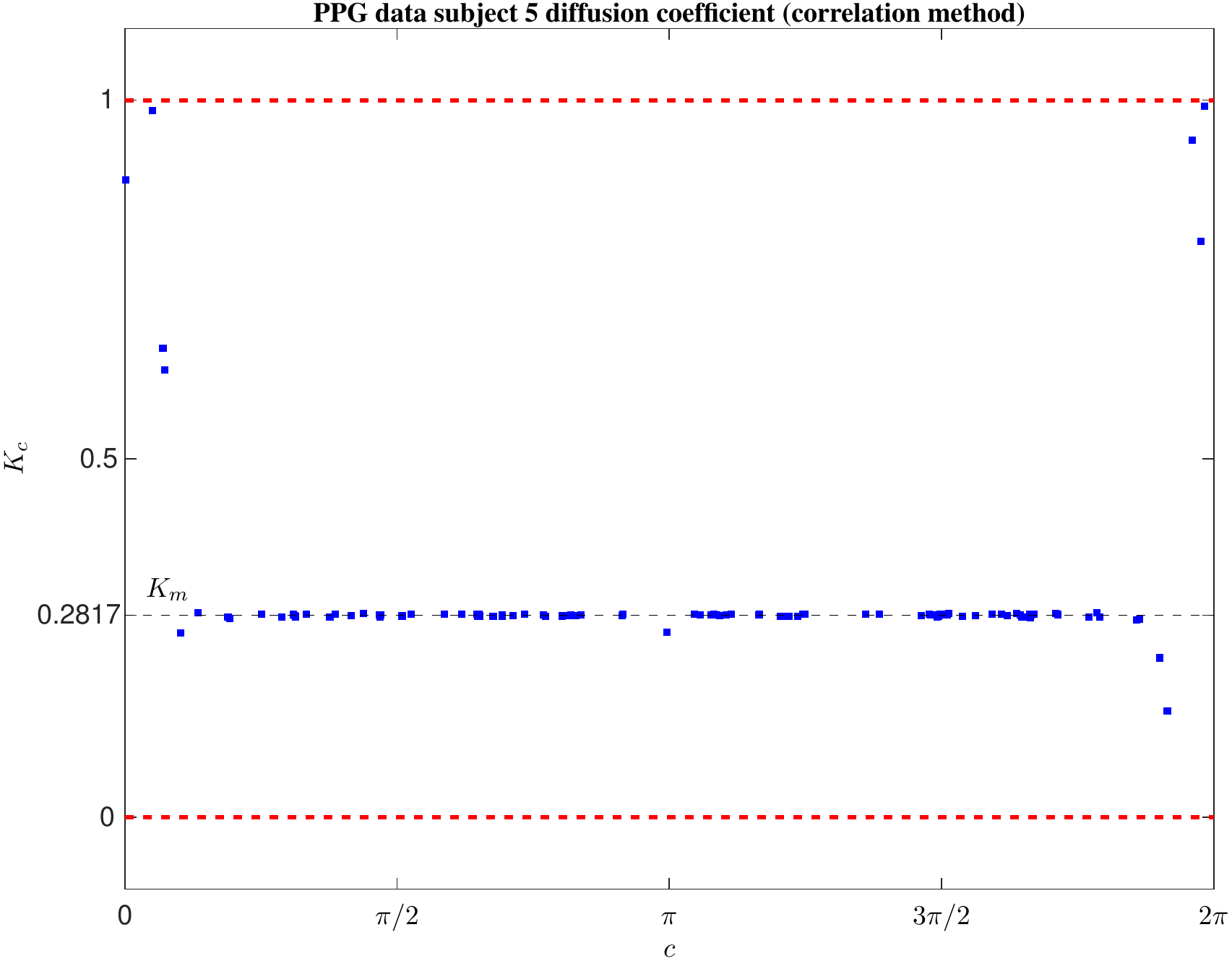}
\label{fig_ten_case}}
\caption{Each column correspond with a subject; number of points: 5,000. (a)-(e) PPG signals, normalized for five subjects; (f)-(j) Representation of equation \eqref{eqn:SD01} for $c=2.5$ for each subject; (k)-(o) $K_{c}$ results of  0--1 test, for 100 random $c$ values, with correlation method.}
\label{fig:F02}
\end{figure*}
To study the dynamical system that represents the signals just looking to a graph, the representation $p_{c}(n)$ and $q_{c}(n)$ in Fig. \ref{fig:F01}(f)-(j) shows perfectly how the periodic signal describes a precise orbit and the aperiodic covers all the space in a restriction area; as opposed to the chaotic and random signals that tend to expand. The quasi-periodic signal presents a behavior between periodic and aperiodic. A way to differentiate chaotic and random behavior is to evaluate the velocity to spread,  checking values of the coordinates of the graph. 

The most important results are the ones of Fig. \ref{fig:F01}(k)-(o) where the values of $K_{m}$ for a periodic signal (in the case of a sine wave, at $f=100$ Hz, is $0.177$) is $0.094$; quasi-periodic is $0.357$; aperiodic is $0.520$; chaotic is $\approx1$, and random is $\approx1$.  

In Fig. \ref{fig:F02} we show results of the PPG signal for five subjects choose randomly from the 40 students, between 18 and 30 years old and a non-regular consumer of psychotropic substances, alcohol or tobacco, according to project details \cite{Aguilo2015}. In this case, it is shown directly the temporal PPG signal used for the study. The PSD graphs will not give significant information about the signal. Remember PPG fundamental frequency, typically around 1 Hz, depending on heart rate (0.5--4 Hz principal and first harmonic); and respiratory activity in 0.2--0.35 Hz \cite{Ram2012}. In our case, all PPG signals have been filtrated to avoid motion artifacts as much as possible. The graphs shown for all the subjects in Fig. \ref{fig:F02}(f)-(j) are quite similar, and if we look and compare to Fig. \ref{fig:F01}, it is evident that the PPG signal independently of the healthy young subject is similar to a quasi-periodic signal. Further, if the 0--1 test is applied, we observe in Fig. \ref{fig:F02}(k)-(o) that for all of them the value of $K_{m}$ is less than $0.5$ and bigger than $0.2$, in such a way that we can affirm that the PPG signal in healthy young people has a quasi-periodic respond; as it has been told for all biological signals in several publications but not with a clear definition of quasi-periodicity just as refereed in \cite{Ganeshapillai2012}.

\begin{figure*}[!t]
\centering
\subfloat[]{\includegraphics[width=1.25in]{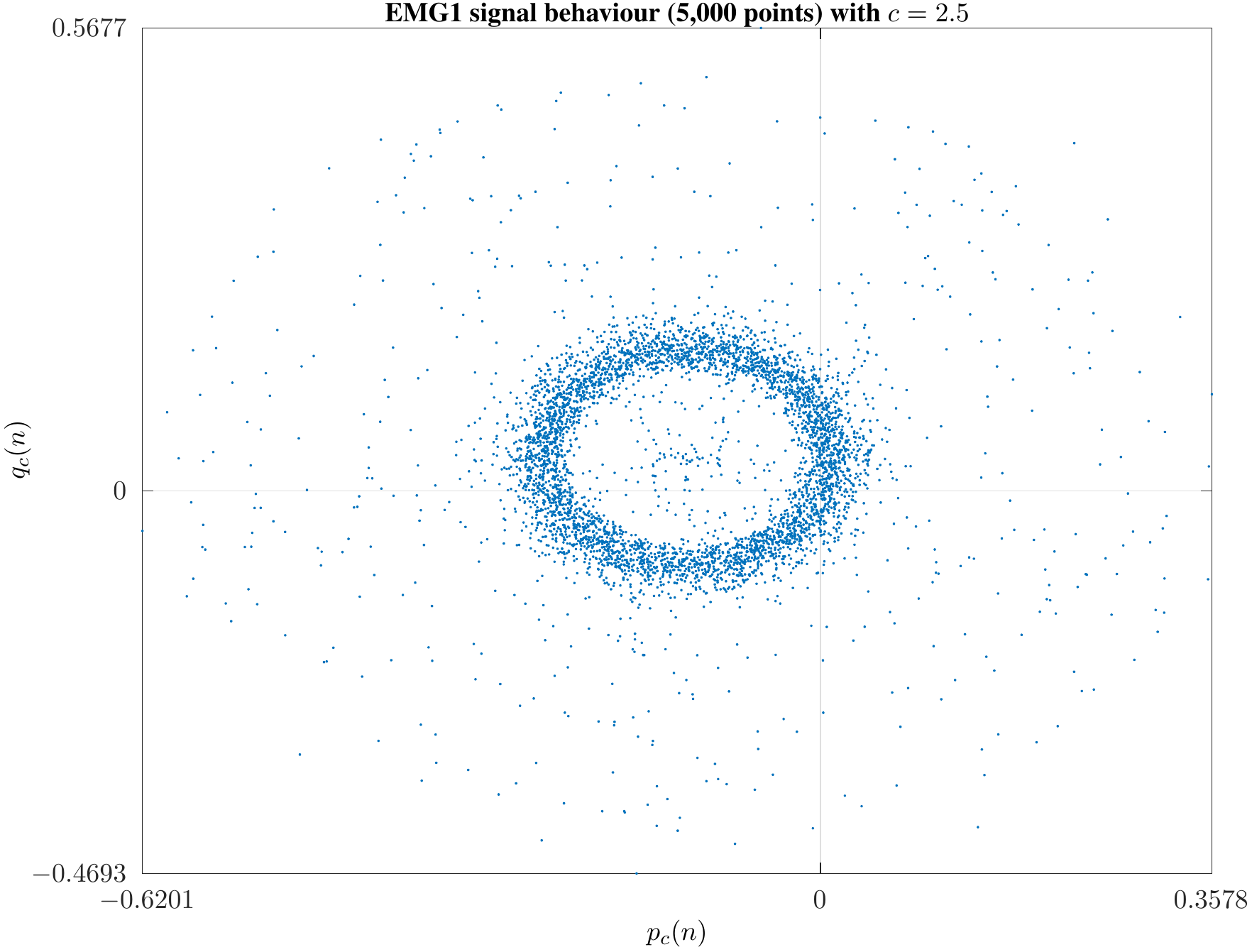}
\label{fig_six_case}}
\subfloat[]{\includegraphics[width=1.25in]{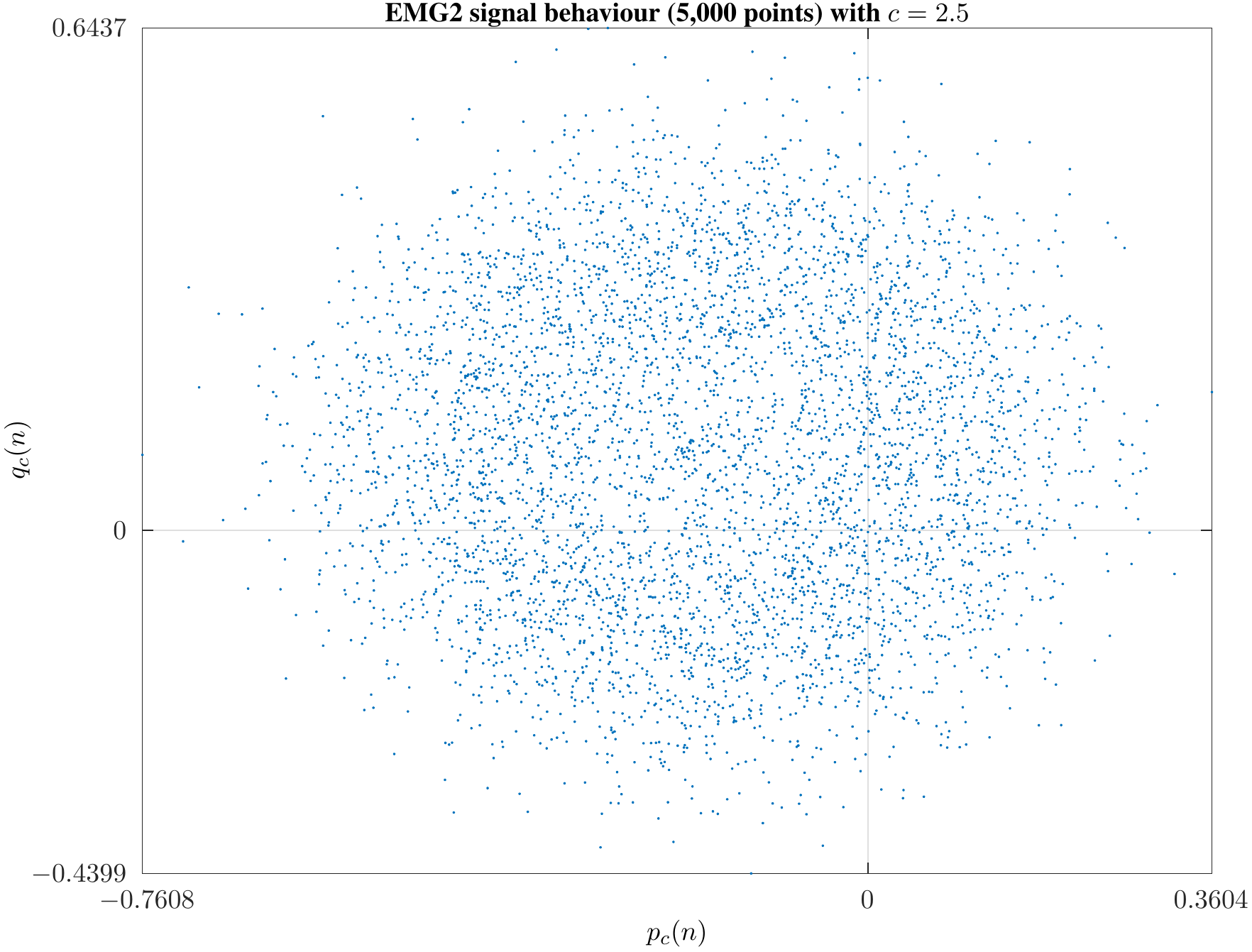}
\label{fig_seven_case}}
\subfloat[]{\includegraphics[width=1.25in]{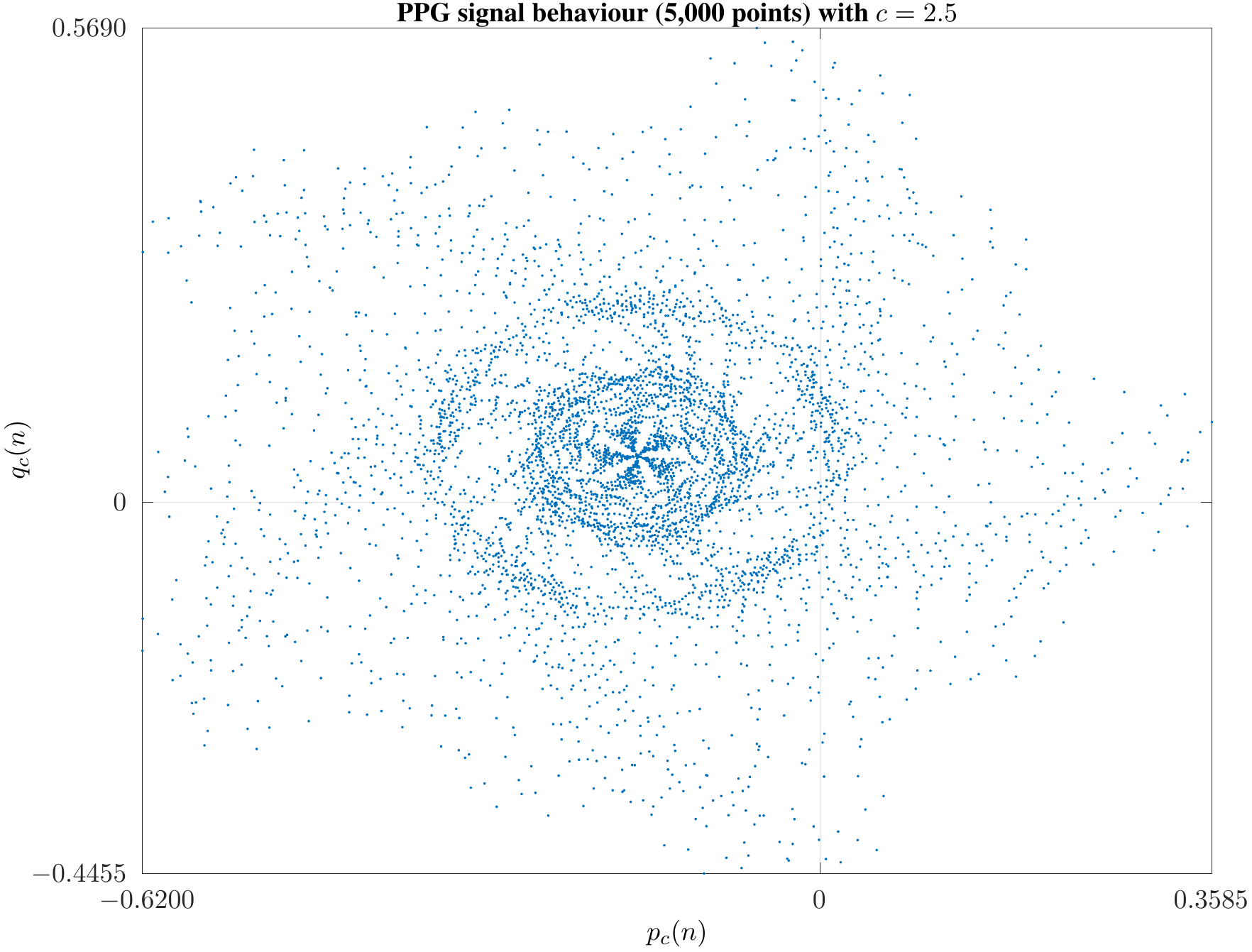}
\label{fig_eight_case}}
\subfloat[]{\includegraphics[width=1.25in]{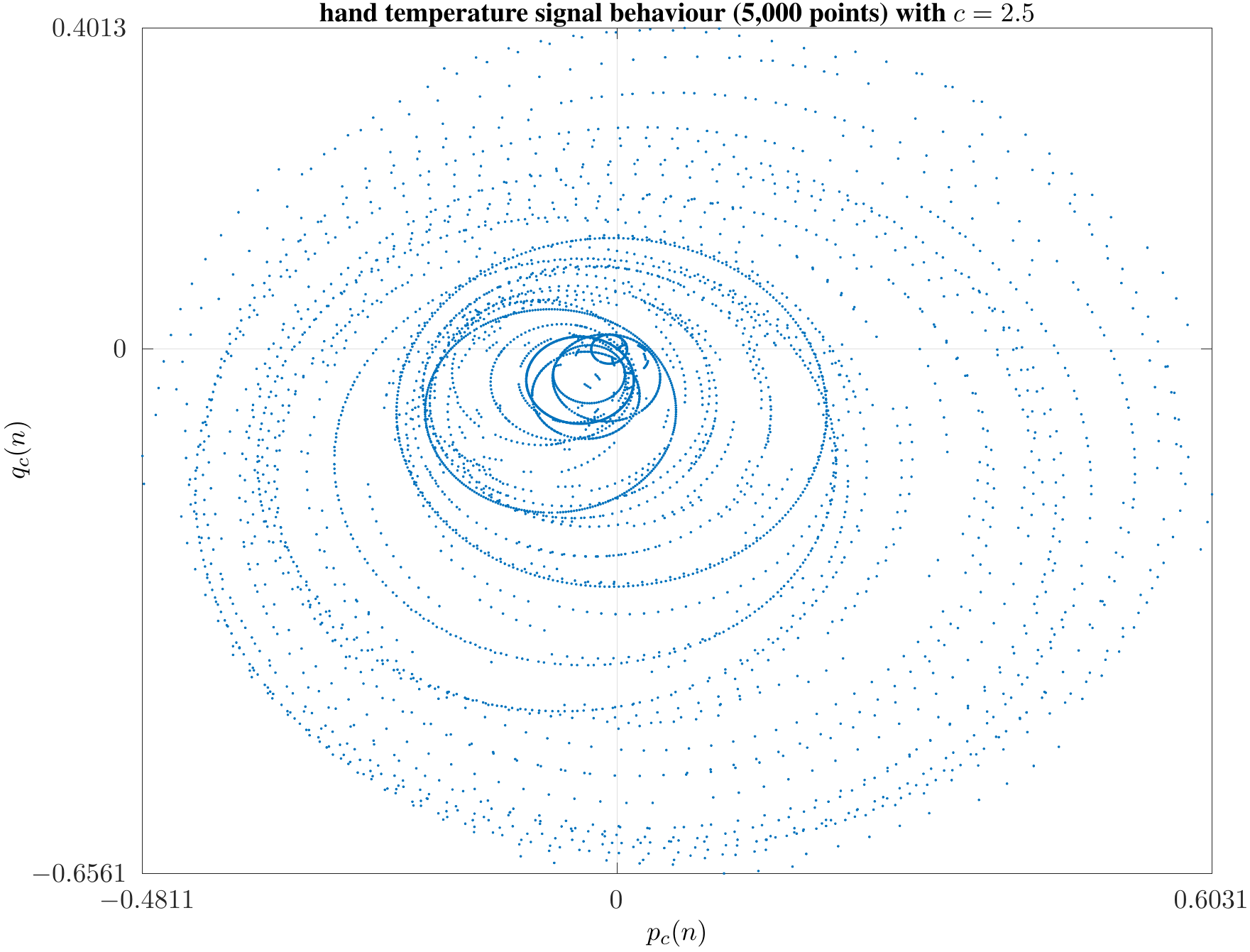}
\label{fig_nine_case}}
\subfloat[]{\includegraphics[width=1.25in]{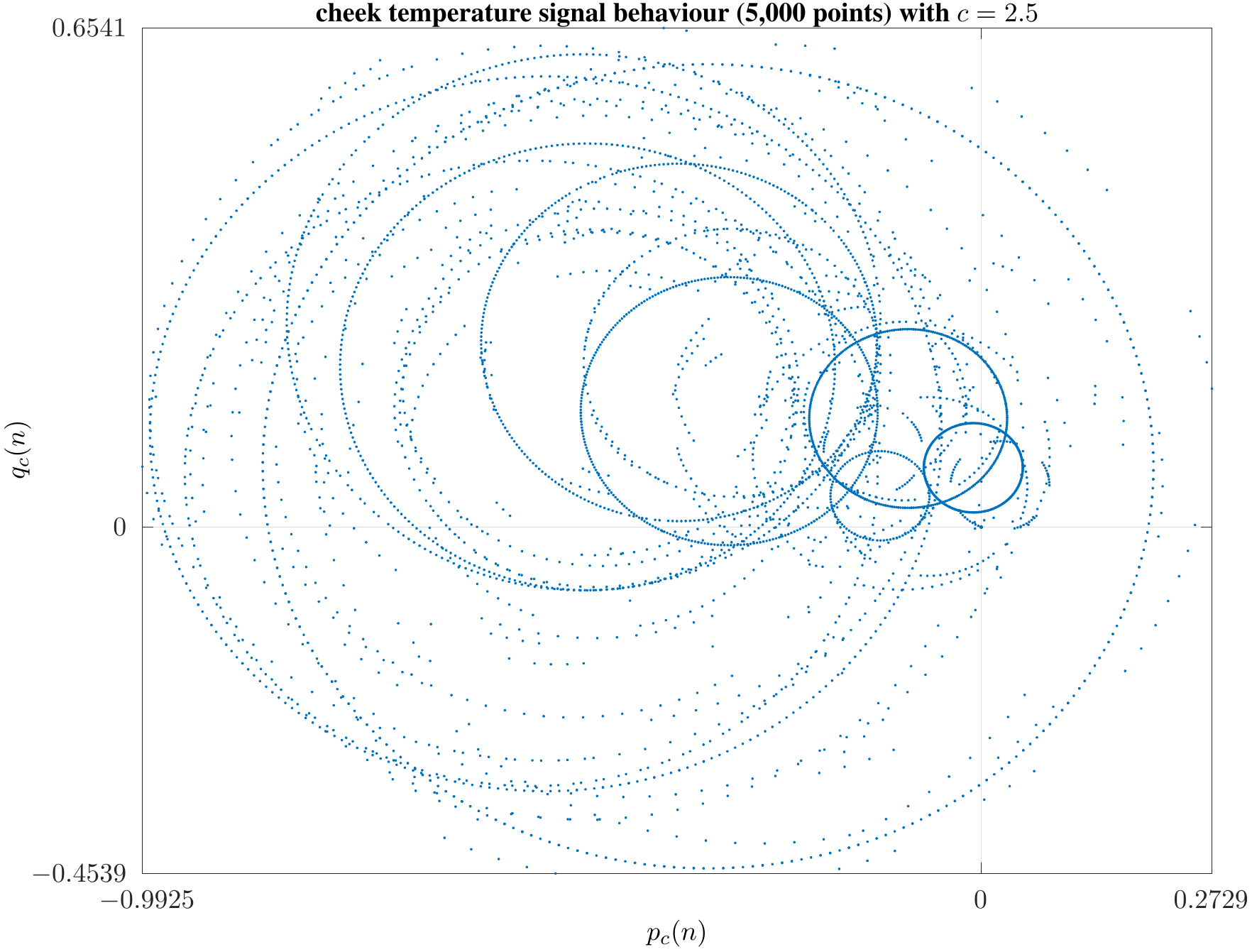}
\label{fig_ten_case}}
\hfil
\subfloat[]{\includegraphics[width=1.25in]{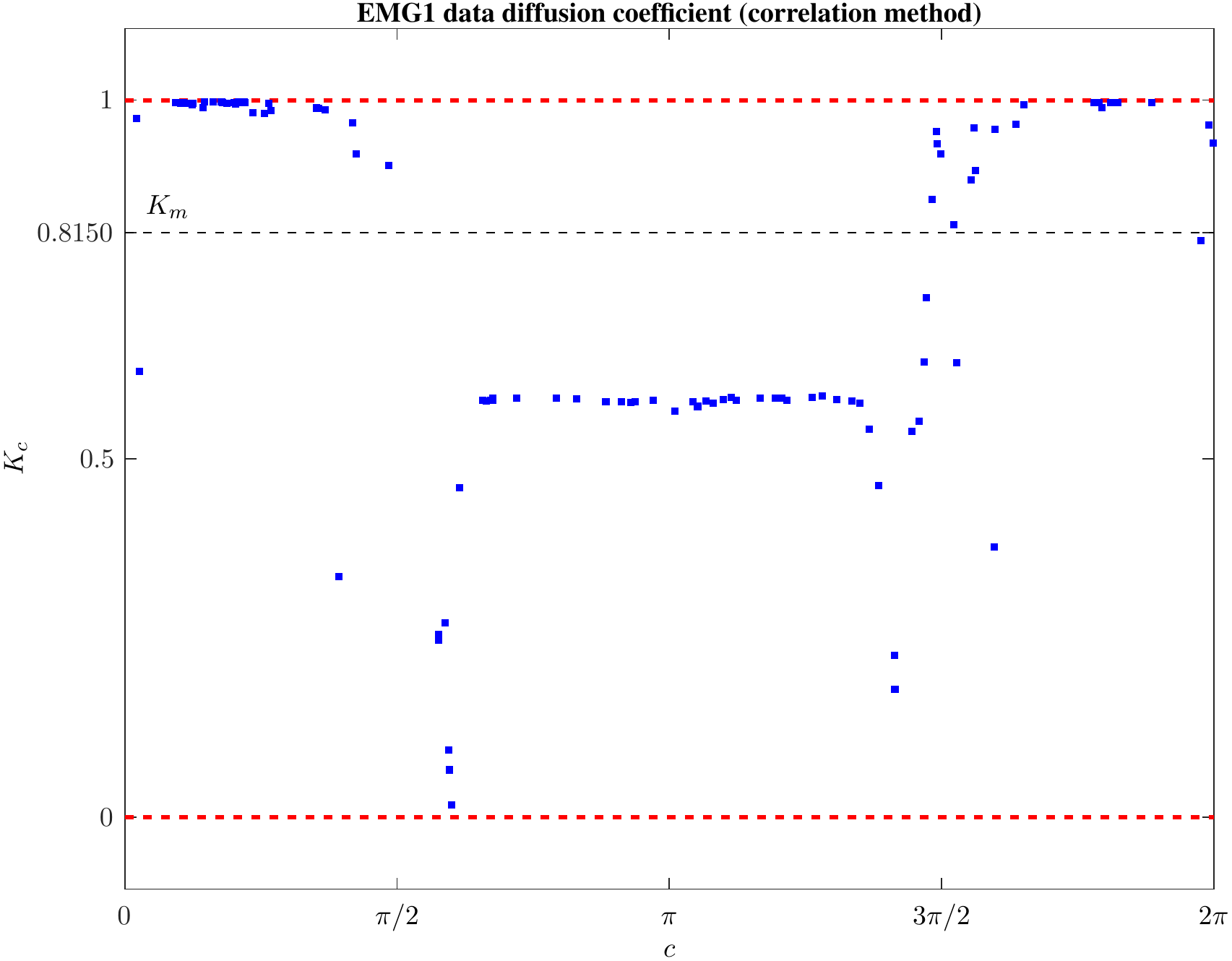}
\label{fig_six_case}}
\subfloat[]{\includegraphics[width=1.25in]{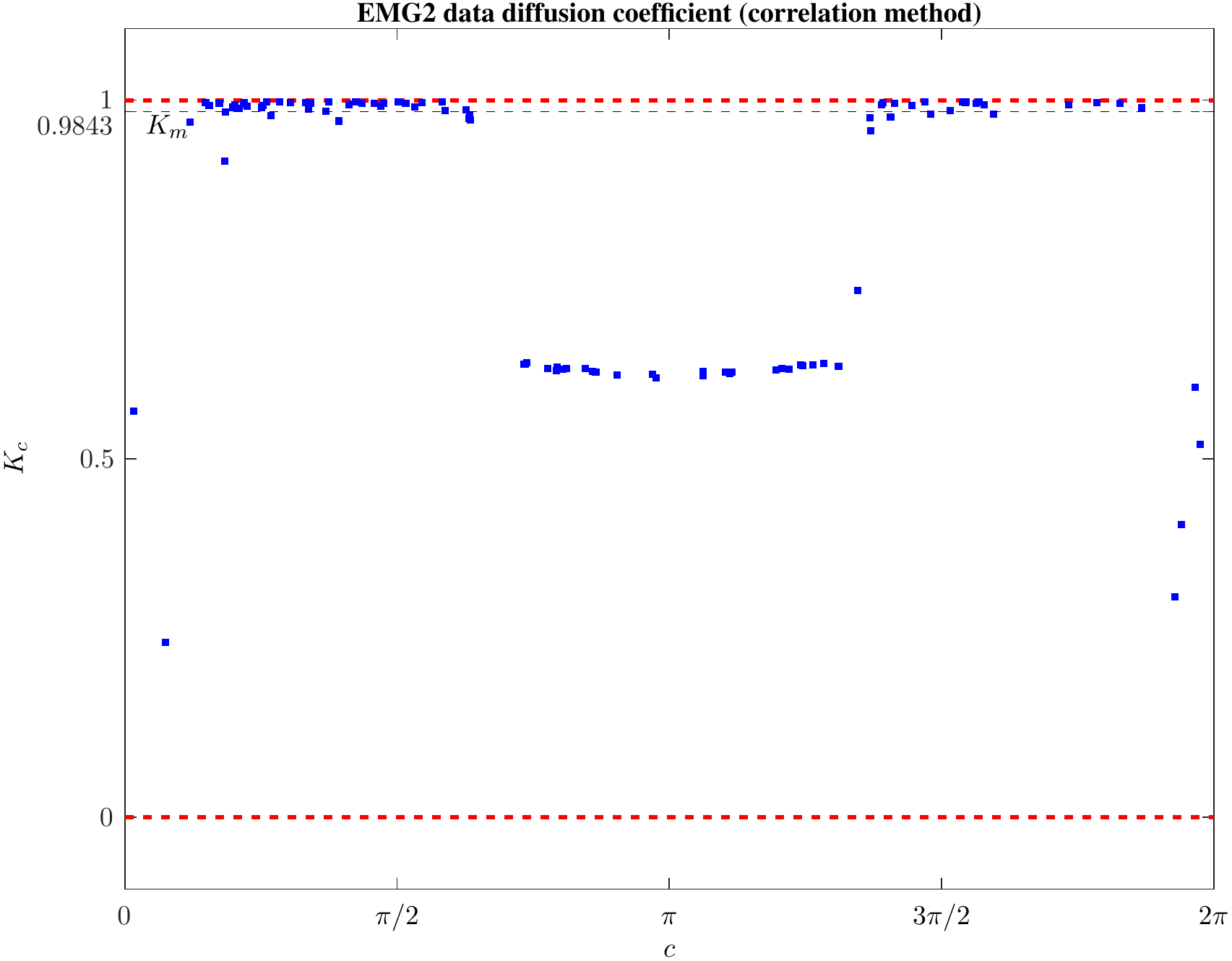}
\label{fig_seven_case}}
\subfloat[]{\includegraphics[width=1.25in]{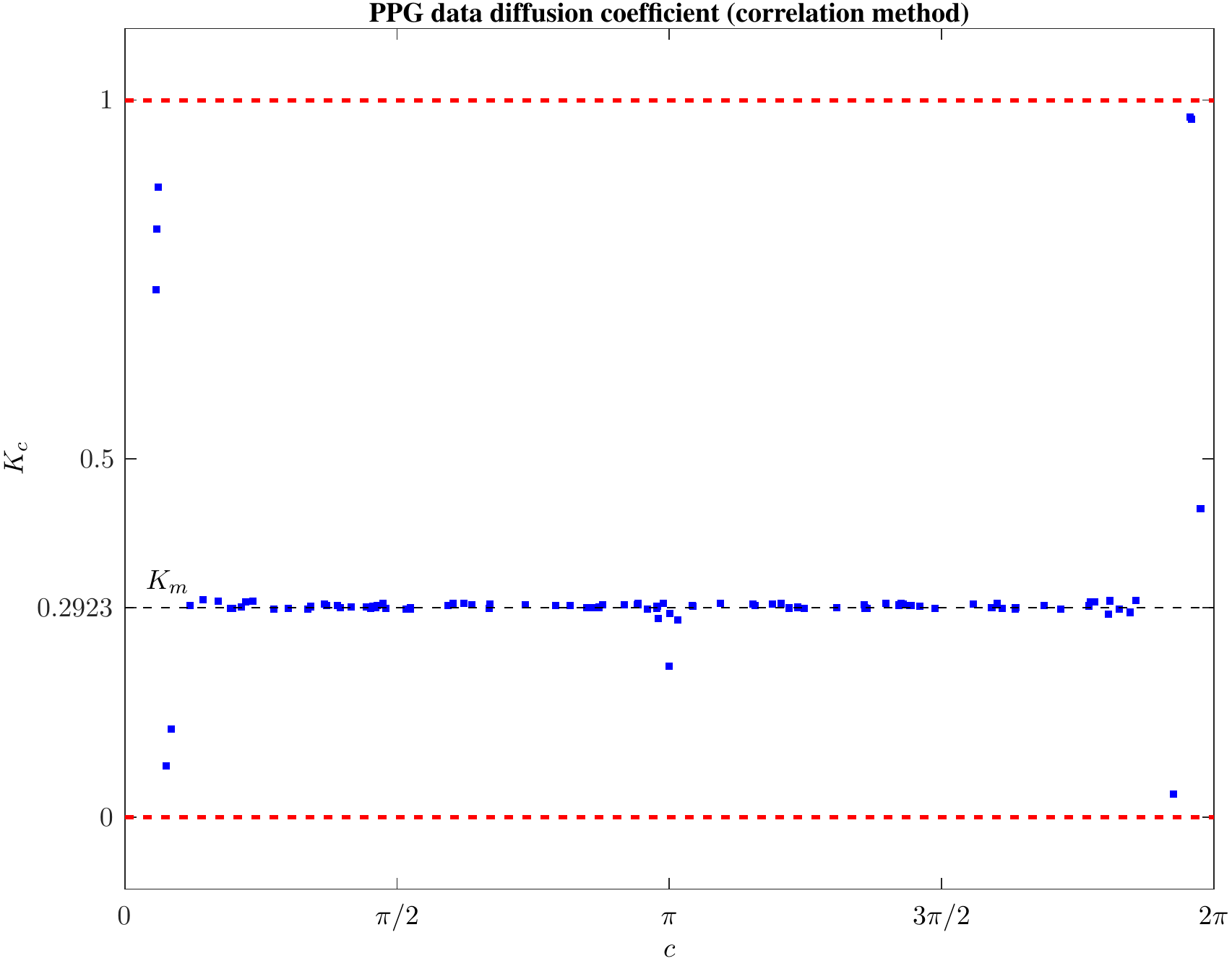}
\label{fig_eight_case}}
\subfloat[]{\includegraphics[width=1.25in]{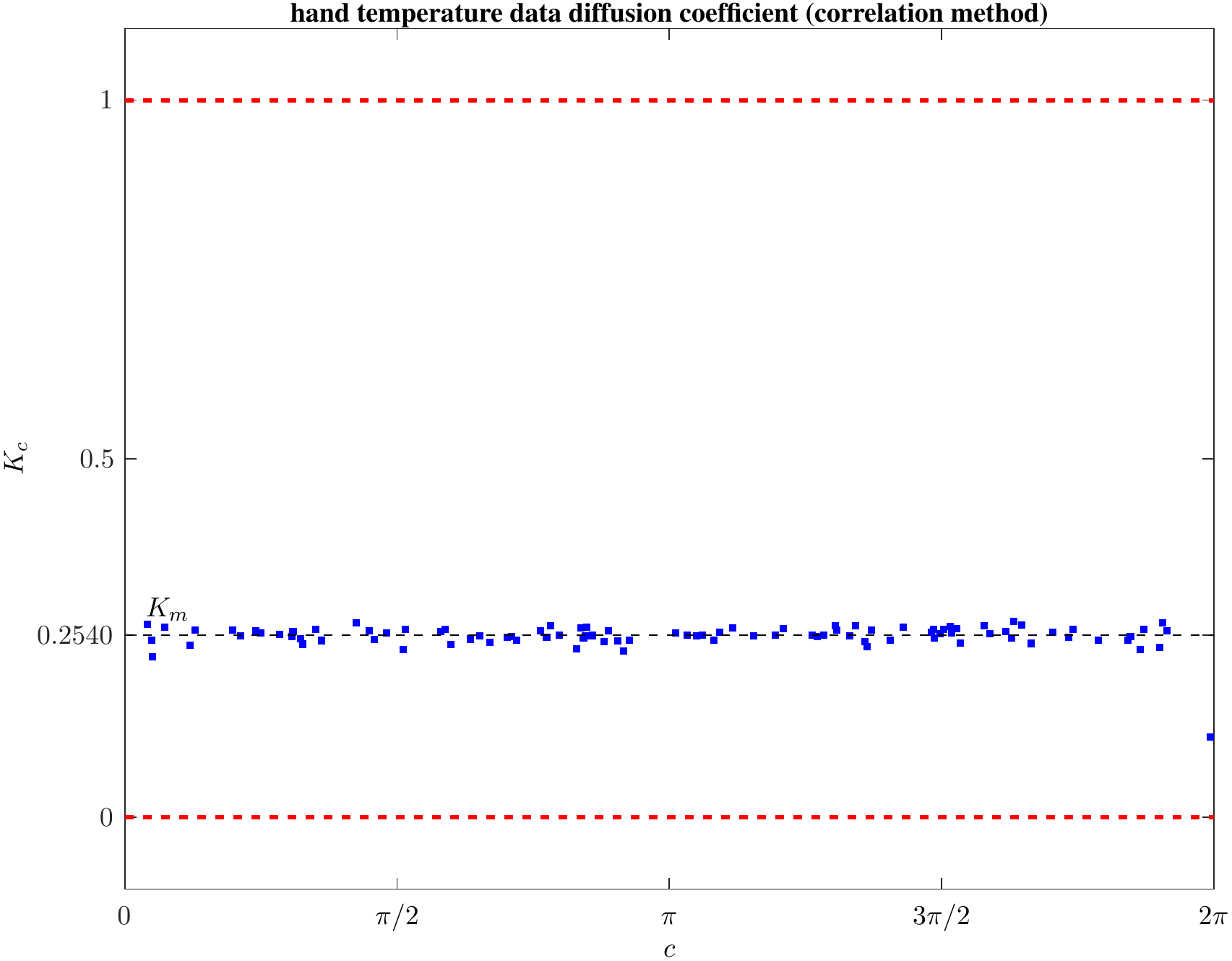}
\label{fig_nine_case}}
\subfloat[]{\includegraphics[width=1.25in]{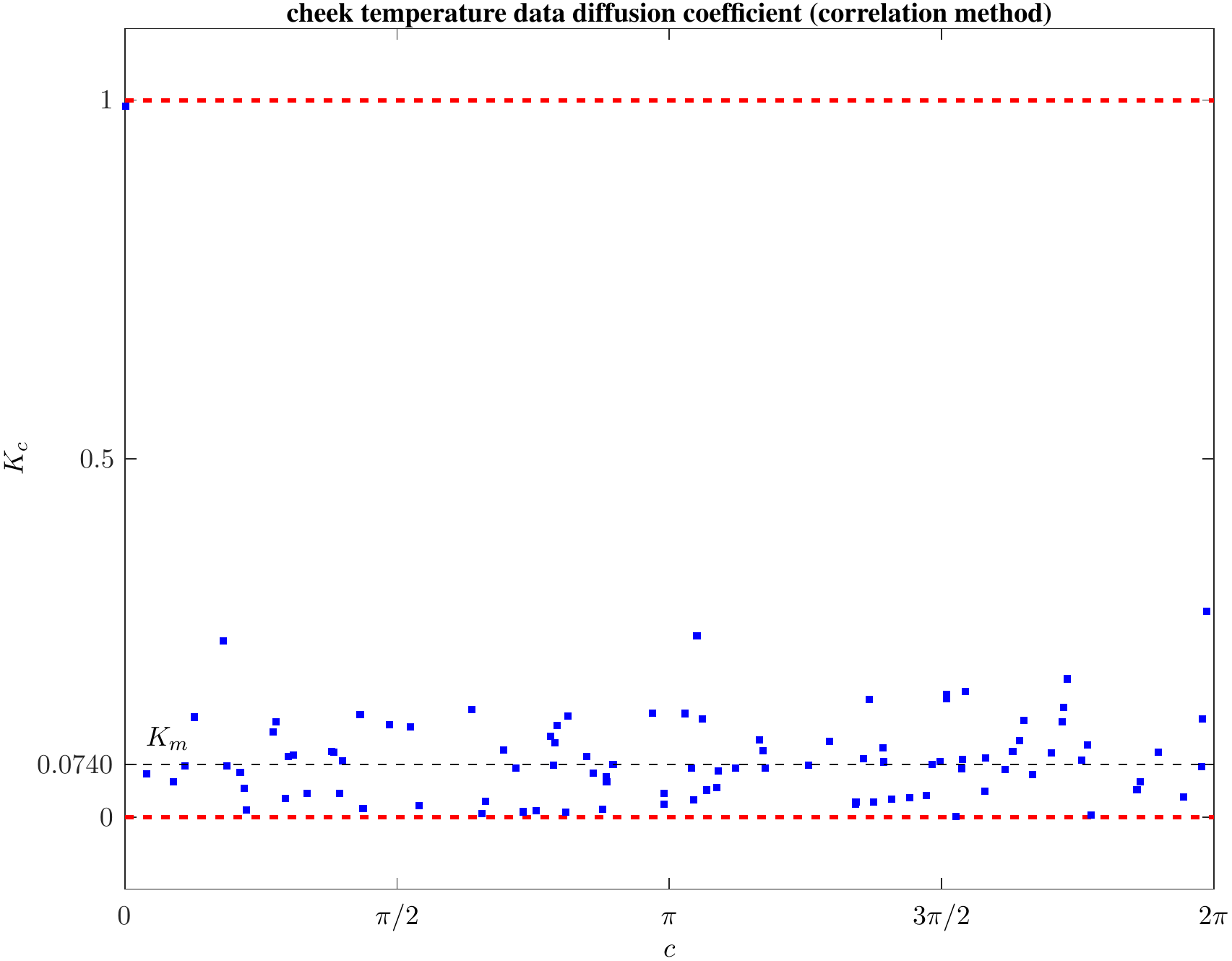}
\label{fig_ten_case}}

\caption{From the same subject, each column correspond with a: EMG signal in trapezius muscle and face (under the eye), PPG signal, finger and cheek temperature; number of points: 5,000. (a)-(e) Representation of equation \eqref{eqn:SD01} for $c=2.5$ for each subject; (k)-(o) $K_{c}$ results of  0--1 test, for 100 random $c$ values, with correlation method.}
\label{fig:F03}
\end{figure*}

Finally Fig. \ref{fig:F03} shows the results of applying the 0--1 test to different biological signals, included the PPG signal, of a subject.  In two first columns of Fig. \ref{fig:F03}, (a)-(f) and (b)-(g), we see that, on the representation of equation \eqref{eqn:SD01} for  Electromyography (EMG), the behavior for the two points where the data is measured is not equal; for further information we need to obtain the values of $K_{c}$. In Fig. \ref{fig:F03}(f)-(g), we see different areas on its values; the reason is that there are some problems with the measurement of EMG signals. In only two seconds, 5,000 points, it detects that it is an area with a tendency to chaotic or random behavior, sees Fig. \ref{fig:F01}(n)-(o), and two regions with a complete chaotic or random behavior. It seems that the noise amplitude present is not constant; something is wrong on the signal measurement---effectively, the EMG sensors were working wrong.

With respect the finger and cheek temperature, the two last columns of Fig. \ref{fig:F03}, (d)-(e) and (i)-(j), it demonstrates that each temperature follows a complete different behavior. Further study of finger and cheek temperature must be done.

\section{Conclusion}\label{sec:CONCLUSION}

We demonstrate that the PPG signal in healthy young subject corresponds to a quasi-periodic behavior of the human body. When a PPG signal has an approximation to chaotic behavior seems that measurement noise is present or maybe some disturbing body respond is present, as can be the stress. Future works will respond to this aspect.

We believe that the dynamics of a PPG signal from a healthy individual respond to a quasi-periodic behavior resulting from the mutual coupling between the two predominant subsystems in the cardiovascular system. These subsystems regulate the peripheral blood flow, that is, cardiac and respiratory activities. The respiratory activity modulates the cardiac regularity imposed by the heartbeat, whose frequency keeps an irrational relation with the cardiac rate. Changes in respiratory and cardiac amplitude and frequency, as a consequence of, for example, a stressful episode, will alter the dynamic configuration of the PPG signal and will lead to aperiodic or chaotic behavior. In this way, the organism would have more modes of operation with which to quickly restore the healthy operating regime.

\section*{Acknowledgment}
The authors would like to thanks to Life Supporting Technologies Group (LST-UPM) for taking part in project FIS-PI12/00514, from MINECO.

\section*{References}

\bibliographystyle{elsarticle-num}
\bibliography{PPG_signal_dynamical_analysis}

\end{document}